\documentclass{aa}  

\usepackage{graphicx}
\usepackage{txfonts}
\usepackage{threeparttable}
\usepackage{multirow}
\begin{document}

   \title{Multi-wavelength properties of three new radio-powerful $z\sim5.6$ quasi-stellar objects discovered from RACS}

   \author{
L. Ighina\inst{1}
\and
A. Caccianiga\inst{1}
\and
A. Moretti\inst{1}
\and
J. W. Broderick\inst{2,3}
\and
J. K. Leung\inst{4,5,6}
\and 
A. R. L\'opez-S\'anchez\inst{7,8,9}
\and
F. Rigamonti\inst{10,11}
\and 
N. Seymour\inst{12}
\and
T. An\inst{13,14,15}
\and
S. Belladitta\inst{16,17} 
\and
S. Bisogni\inst{18}
\and
R. Della Ceca\inst{1}
\and
G. Drouart\inst{12}
\and
A. Gargiulo\inst{18}
\and
Y. Liu\inst{13}
          }

\institute{INAF, Osservatorio Astronomico di Brera, via Brera 28, 20121, Milano, Italy \\
\email{ luca.ighina@inaf.it}
         \and
SKA Observatory, Science Operations Centre, CSIRO ARRC, 26 Dick Perry Avenue, Kensington, WA 6151, Australia 
        \and
CSIRO Space \& Astronomy, PO Box 1130, Bentley, WA 6102, Australia
        \and
David A. Dunlap Department of Astronomy \& Astrophysics, University of Toronto, 50 St. George St., Toronto, Ontario, M5S 3H4, Canada
        \and
Dunlap Institute for Astronomy \& Astrophysics, University of Toronto, 50 St. George Street Toronto, ON M5S 3H4, Canada
        \and
Racah Institute of Physics, The Hebrew University of Jerusalem, Jerusalem, 91904, Israel
        \and
Australian Astronomical Observatory, 105 Delhi Road, North Ryde, NSW 2113, Australia
        \and
Department of Physics and Astronomy, Macquarie University, NSW 2109, Australia
        \and
Australian Research Council Centre of Excellence for All-Sky Astrophysics in 3 Dimensions (ASTRO 3D), Australia
        \and
INFN, Sezione di Milano-Bicocca, Piazza della Scienza 3, I-20126 Milano, Italy
        \and
INAF – Osservatorio Astronomico di Brera, via E. Bianchi 46, I–23807 Merate, Italy
        \and
International Centre for Radio Astronomy Research, Curtin University, 1 Turner Avenue, Bentley, WA, 6102, Australia
        \and
Shanghai Astronomical Observatory, Chinese Academy of Sciences (CAS), 80 Nandan Road, Shanghai 200030, China
        \and
School of Astronomy and Space Sciences, University of Chinese Academy of Sciences, No. 19A Yuquan Road, Beijing 100049, China
        \and
Key Laboratory of Radio Astronomy and Technology, CAS, A20 Datun Road, Beijing, 100101, P. R. China
        \and
Max Planck Institut für Astronomie, Königstuhl 17, D-69117, Heidelberg, Germany
        \and
INAF – Osservatorio di Astrofisica e Scienza dello Spazio di Bologna, Via Gobetti 93/3, I-40129 Bologna, Italy
        \and
INAF - Istituto di Astrofisica Spaziale e Fisica Cosmica (IASF), Via A. Corti 12, 20133, Milano, Italy        }

   \date{Received September 15, 1996; accepted March 16, 1997}

 
  \abstract{We present a multi-wavelength study of three new $z\sim5.6$ quasi-stellar objects (QSOs) identified from dedicated spectroscopic observations. The three sources were selected as high-$z$ candidates based on their radio and optical/near-infrared properties as reported in the Rapid ASKAP Continuum Survey (RACS), the Dark Energy Survey (DES), and the Panoramic Survey Telescope and Rapid Response System (Pan-STARRS) survey. These are among the most radio-bright QSOs currently known at $z>5.5$, relative to their optical luminosity, having $\rm R=S_{\rm 5GHz}/S_{\rm 4400\AA}>100$. In this work, we present their identification, and we also discuss their multi-wavelength properties (from the radio to the X-ray band) based on detections in public surveys as well as in dedicated radio and X-ray observations. The three sources present a wide range of properties in terms of relative intensity and spectral shape, highlighting the importance of multi-wavelength observations in accurately characterising these high-$z$ objects. In particular, from our analysis we found one source at $z=5.61$ that presents clear blazar properties (strong radio and X-ray emission), making it one of the most distant currently known in this class. Moreover, from the fit of the optical/near-infrared photometric measurements with an accretion disc model as well as the analysis of the CIV broad emission line in one case, we were able to estimate the mass and accretion rate of the central black holes in these systems, finding $\rm M_{\rm BH}\sim1-10\times10^9$~M$_\odot$ accreting at a rate $\lambda_{\rm Edd}\sim0.1-0.4$. The multi-wavelength characterisation of radio QSOs at $z>5.5$, such as the ones reported here, is essential to constraining the evolution of relativistic jets and supermassive black holes hosted in this class of objects.  
}

 \titlerunning{Three new $z\sim5.6$ radio-powerful QSOs}

   \maketitle
%
 
\section{Introduction}

One of the main unsolved questions in current astrophysics and cosmology is how the most massive supermassive black holes (SMBHs) in the early Universe ($z\gtrsim6$) formed and grew into the large masses currently observed ($\sim10^9~{\rm M_{\odot}}$; \citealt{Banados2018,Wang2021,Fan2023}) in such a short time \citep[e.g.][]{Volonteri2021}.
Two of the main ingredients to build up massive black holes (BHs) in the early Universe are the initial seed BH \citep[e.g.][]{Pezzulli2017,Valiante2016,Singh2023} and the accretion rate at which these seeds grow \citep[e.g.][]{Johnson2013,Pacucci2015, Lupi2024b}. Therefore, the two most common explanations for the presence of $>10^9~{\rm M_\odot}$ SMBHs at $z\sim6$ involve an already massive seed BH at $z\sim10-15$ \citep[e.g.][]{Begelman2006,Lu2024} and/or a sustained accretion rate over the Eddington limit \citep[e.g.][]{Johnson2022,Lupi2024}. Both of these scenarios have been put to the test with the observations of active galactic nuclei hosting SMBHs with masses from $>10^{6-7}~{\rm M_\odot}$ \citep[e.g.][]{Larson2023,Kokorev2023,Maiolino2024} up to $z\sim8-10$ likely right after their formation ($<0.3$~Gyr).

However, the jetted quasi-stellar object (QSO) population may also offer a viable solution to explain the large masses that are observed. Indeed, if, besides radiation, part of the gravitational energy released by the accreting matter is converted into magnetic and/or mechanical energy (e.g. to power relativistic jets; \citealt{Jolley2008,Jolley2009}), these systems can actually accrete more matter compared to QSOs with a similar optical luminosity but without jets (see, e.g., \citealt{Connor2024}).

One way to trace the evolution of the entire jetted QSO population is to use the specific class of blazars \citep[e.g.][]{Sbarrato2022,Diana2022}. These are a particular subset of radio-loud (RL\footnote{From an observational point of view, a QSO is defined to be radio loud when its radio loudness parameter is $\rm R=S_{\rm 5GHz}/S_{\rm 4400\AA}>10$ (in the rest frame; \citealt{Kellerman1989}).}) QSOs for which the relativistic jet is oriented close to our line of sight ($\theta_{\rm v}\lesssim1/\Gamma$, where $\Gamma$ is the bulk Lorentz factor of the jet). Indeed, from the detection of even a single blazar, one can infer the presence of $\sim2\Gamma^2$ (with $\Gamma\sim10$; e.g. \citealt{Marcotulli2020}) objects with similar properties and at the same redshift but with jets oriented in a different direction (assuming an isotropic distribution of orientations; e.g. \citealt{Caccianiga2024,Banados2024}) and therefore much lower observed radio luminosities, even if they are obscured (e.g. \citealt{Endsley2022,Endsley2023,Lambrides2024}). 

However, current studies focused on blazars are limited by low statistics. Indeed, at the start of the work presented here, only two QSOs were classified as blazars at $z>5.5$: HZQ~J0901+16 at $z=5.63$ \citealt{Caccianiga2024} and HZQ~J0309+27 at $z=6.10$ \citealt{Belladitta2020}. 
While several efforts are being made to increase the statistics for high-$z$ RL QSOs through the combination of radio and optical/near-infrared (NIR) surveys \citep[e.g.][]{Gloudemans2022,Ighina2023,Caccianiga2024}, the number of radio sources with dedicated X-ray information is still limited \citep[e.g.][]{Medvedev2021,Wolf2021,Khorunzhev2021,Zuo2024}.

In this work we present the discovery and properties of three radio-bright QSOs at redshift $z\sim5.6$ selected from the combination of the first data release of the Rapid Australian SKA Pathfinder Telescope (ASKAP) Continuum Survey (RACS; \citealt{McConnell2020,Hale2021}) in the radio band (centred at 888~MHz) with the Panoramic Survey Telescope and Rapid Response System \citep[Pan-STARRS; ][]{Chambers2016} and the Dark Energy Survey \citep{Abbott2021} in the optical/NIR band. Given their high-redshift and radio-bright nature, we performed dedicated radio and X-ray follow-up observations aimed at constraining the properties of their jets.


Throughout the paper we assume a flat $\Lambda$CDM cosmology with $H_{0}$=70 km~sec$^{-1}$ Mpc$^{-1}$, $\Omega_m$=0.3, and $\Omega_{\Lambda}$=0.7. Spectral indices are given assuming S$_{\nu}\propto \nu^{-\alpha}$, and all errors are reported at a 68\% confidence level, unless otherwise specified.

\section{Selection, identification, and optical data} 
\label{sec:opt_data}
The sources discussed in this work belong to a larger sample selected from the cross-match of the RACS-low radio survey \citep[centred at 888~MHz;][]{McConnell2020} together with DES \citep{Abbott2021} and Pan-STARRS \citep{Chambers2016} optical/NIR surveys by using the Ly$\alpha$ dropout technique (e.g. \citealt{Banados2023}) in the $r-i$ and $i-z$ filters in order to select $z\gtrsim5$ candidates with an apparent magnitude in the $z$-filter \textless21.3 (see, e.g., \citealt{Caccianiga2019,Belladitta2020} for similar studies). The overall sample is composed by 44 high-$z$ candidates identified at the 90\% level. The detailed description of this sample together with its selection criteria and completeness will be presented in a future work (Ighina et al., in prep.). 
In summary, we started by considering all the radio sources reported in the source lists of the RACS-low survey \citep{McConnell2020} with a peak flux density $S_\mathrm{peak}$~\textgreater~1~mJy~beam$^{-1}$ at 888~MHz and with an integrated/peak flux density ratio  $S_\mathrm{int}$/$S_\mathrm{peak}$~\textless~1.5. We then cross-matched these radio sources with the Pan-STARRS and DES optical/NIR catalogues by adopting a radius of 3$''$ and then applied further cuts to their optical/NIR colours, based on the redshift range targeted, in order to select good high-$z$ QSO candidates.
In the next subsection we describe the limits used for the optical and NIR selection of the three high-$z$ QSOs discussed in this paper.

\subsection{Optical selection and photometric measurements}
All the three candidates discussed here were selected using the Ly$\alpha$ dropout technique for sources in the RACS-low survey with a magnitude in the $z$-band $<21.3$. 
In particular, J020228.5$-$170827 (hereafter PSO~J0202$-$17) was selected from Pan-STARRS as an $r$-dropout ($r_{\rm PS1}-i_{\rm PS1}>1.1$), J020916.9$-$562650 (hereafter DES~J0209$-$56)\footnote{We note that the source DES~J0209$-$56 was independently identified by \cite{Wolf2024} based on a selection focused on X-ray sources. Our discovery and multi-wavelength analysis predates this publication, and therefore, in the following we only discuss our dedicated observations.} was selected from DES as an $r$-dropout ($r_{\rm DES}-i_{\rm DES}>1$) and J101155.5$-$013052 (hereafter PSO~J1011$-$01) was selected from Pan-STARRS as an $i$-dropout ($i_{\rm PS1}-z_{\rm PS1}>1.1$).
Furthermore, we also required the objects to be point-like, that is, \texttt{mag\_psf\_i$_{\rm PS1}$--mag\_kron\_i$_{\rm PS1}$\textless0.07} for PSO~J0202$-$17, \texttt{mag\_psf\_z$_{\rm PS1}$--mag\_kron\_z$_{\rm PS1}$\textless0.1} for PSO~J1011$-$01
and \texttt{class\_star\_i$_{\rm DES}$\textgreater0.85} for DES~J0209$-$56. These cuts were chosen in order to maintain a high level of completeness ($\gtrsim$80\%) when selecting $z>5$ QSOs (see, e.g., \citealt{Caccianiga2019}), whereas the choice for different threshold values for the $r$- and $i$-dropouts from Pan-STARRS is related to the fact that images in the $i$-band are deeper with respect to those in the $z$-band, resulting in more reliable magnitude measurements. Similarly to the selection described in \cite{Ighina2023}, we also looked for sources that are faint in the $g$-band. From a practical point of view, this consisted of an upper limit on the $g$-band magnitude reported in the catalogue ($<24$) and a subsequent visual inspection of each image to ensure a true non-detection.
Finally, from the cross-match of the optical/NIR candidates with the mid-infrared (MIR) Wide-field Infrared Survey Explorer catalogue \citep[catWISE;][]{Eisenhardt2020} with a radius $r=2.5''$, we only considered objects with a colour $z_{\rm PS1}-W2<5$ (for Pan-STARRS) or $2<z_{\rm DES}-W2<6$ (for DES) in order to avoid dust-absorbed QSOs at lower redshift (e.g. \citealt{Carnall2015,Caccianiga2017}).

\begin{figure}
        \includegraphics[width=\hsize]{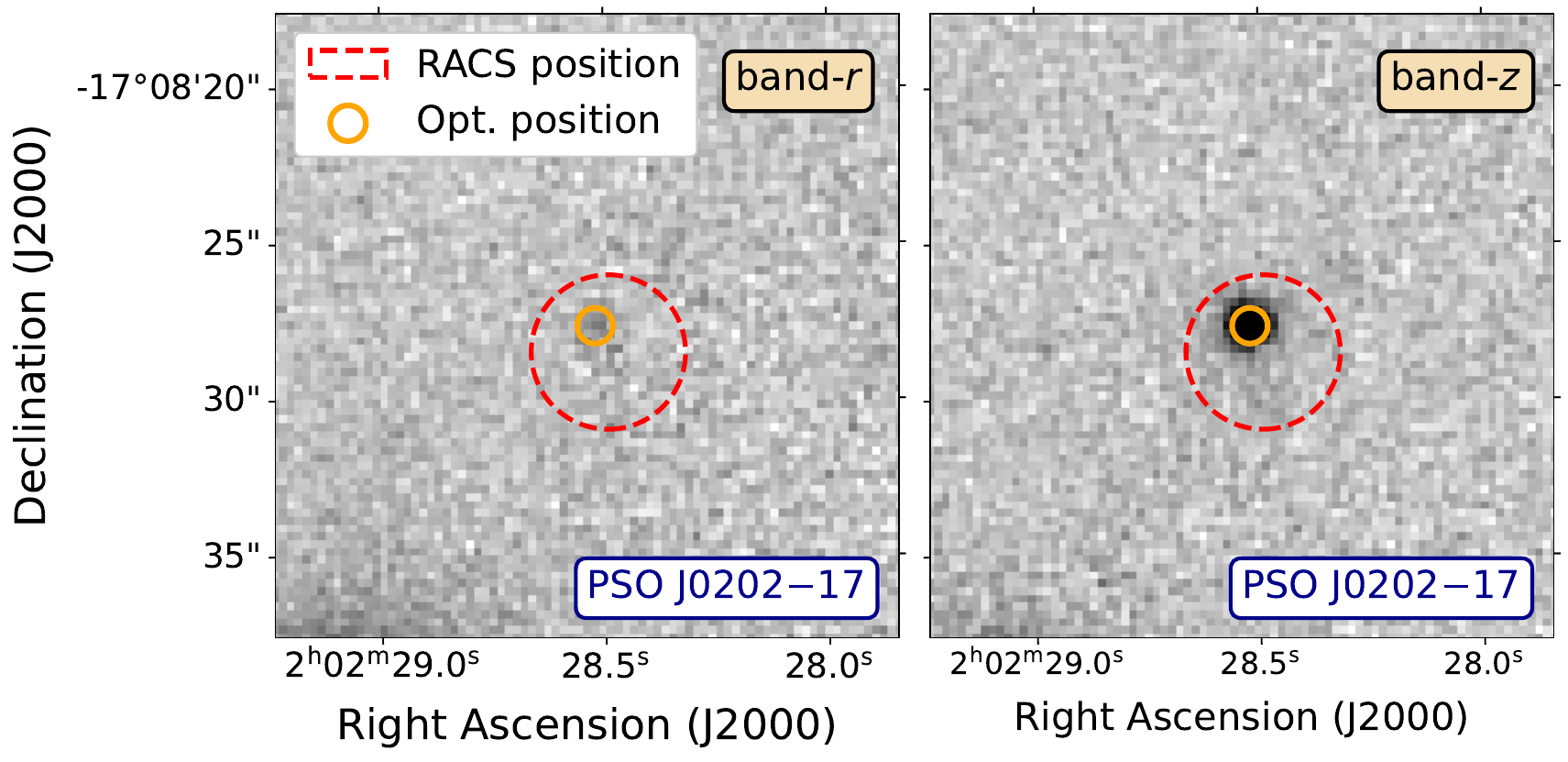}
        \includegraphics[width=\hsize]{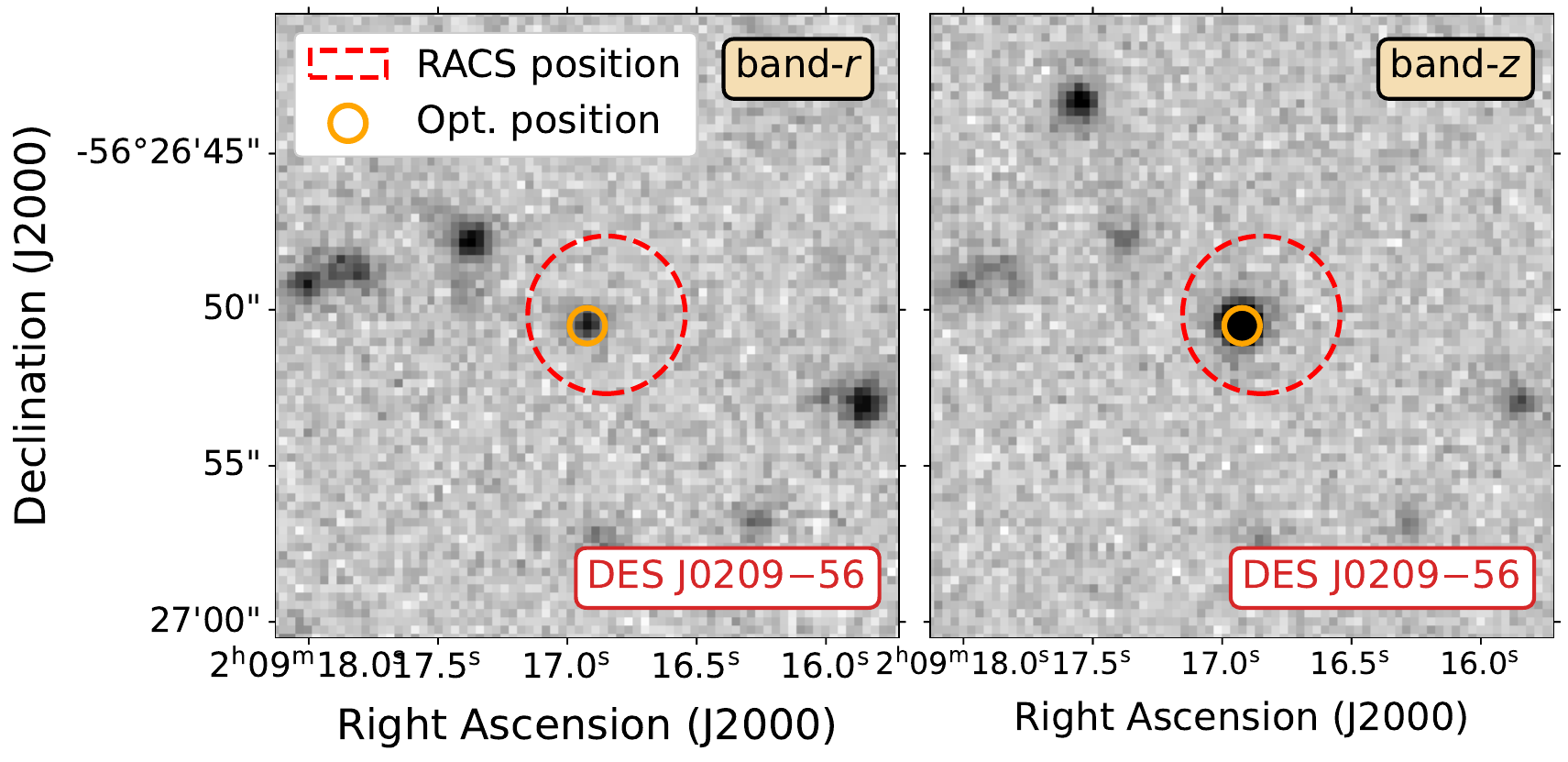}
        \includegraphics[width=\hsize]{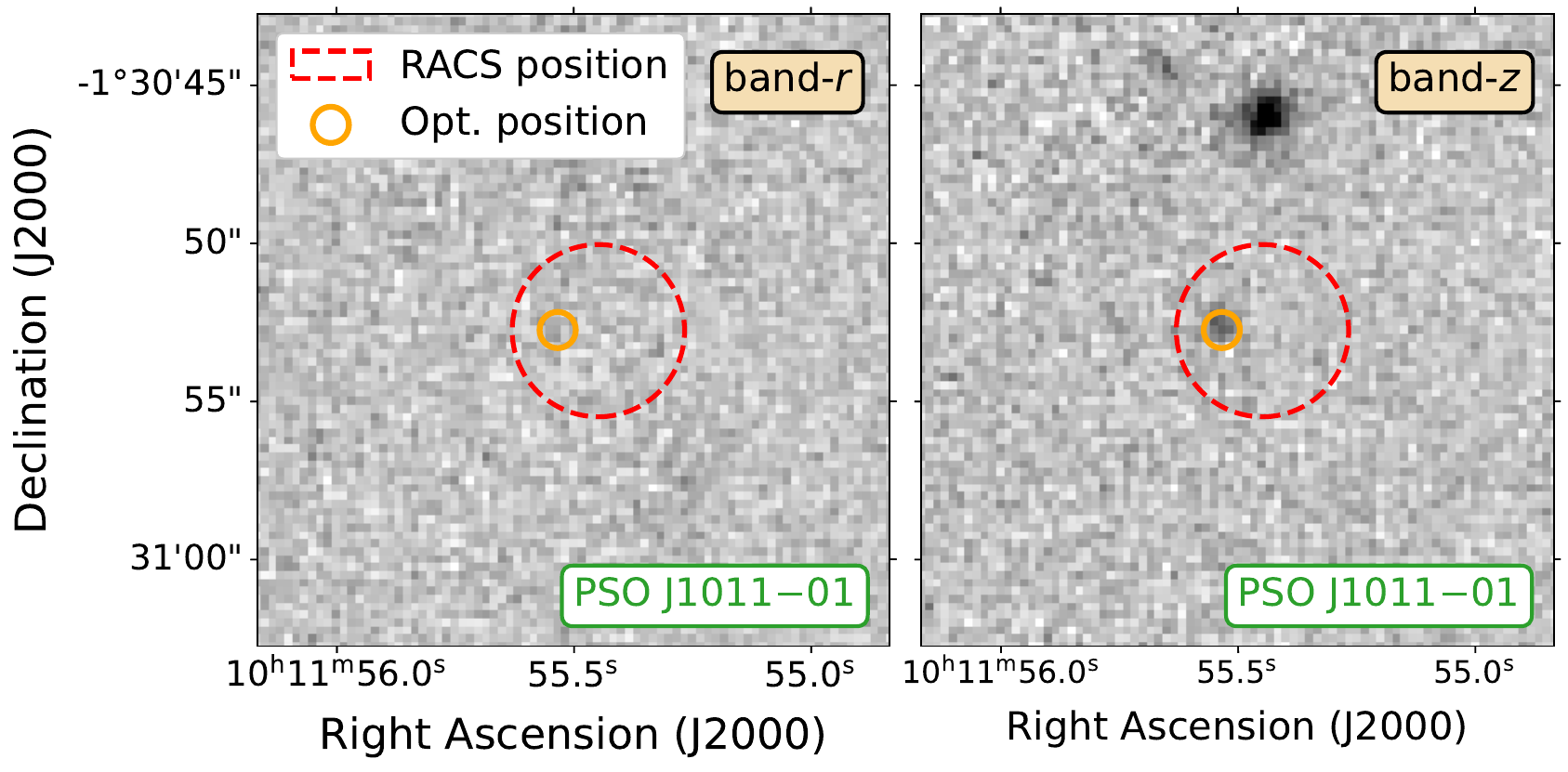}
    \caption{Optical/NIR images ($r$- and $z$-band) of the three sources discussed in this work ($30''\times30''$). Images are from Pan-STARRS in the case of PSO~J0202$-$17 and PSJ1011$-$01 and from DES in the case of DES~J0209$-$56. The solid orange circle indicates the optical position, while the dashed red circle is the uncertainty of the radio position of each object reported the RACS-low source list ($\sim2''$).}
    \label{fig:opt_im}
\end{figure}

In Fig. \ref{fig:opt_im} we show the optical images ($r$ and $z$ band, i.e. before and after the Ly$\alpha$ dropout, respectively) together with the RACS-low position as reported in the source lists derived by \cite{McConnell2020} that we used for the radio-optical association of the candidates. In Table \ref{tab:photom_data} we report the magnitudes of the three targets discussed in this work from different publicly available optical, near- and mid-infrared surveys (all in the AB system). All the magnitudes have been corrected for Galactic absorption assuming $R_{\rm V}=3.1$ \citep{Schlafly2011} and the extinction law from \cite{Fitzpatrick1999}. 

\begin{table*}
    \caption{Magnitudes (AB system) of the three sources discussed in this work from different optical, NIR, and IR surveys.}
    \centering
    \begin{tabular}{cc|ccccc}
        Survey & Filter & & Target & \\
        \hline
        \hline
        & & PSO~J0202$-$17 & DES~J0209$-$56 & PSO~J1011$-$01 \\
        \hline
Pan-STARRS& $r_{\rm PS1}$& 21.95$\pm$0.06 & -- & $<23.20$  \\
        & $i_{\rm PS1}$ & 20.28$\pm$0.04 & -- & 22.31$\pm$0.11 \\
        & $z_{\rm PS1}$ & 19.22$\pm$0.02 & -- & 21.19$\pm$0.08 \\
        & $y_{\rm PS1}$ & 19.07$\pm$0.03 & -- & 21.02$\pm$0.18 \\
    DES & $r_{\rm DES}$ & 21.66$\pm$0.02 & 23.82$\pm$0.11 & -- \\
        & $i_{\rm DES}$ & 19.77$\pm$0.01 & 21.85$\pm$0.03 & --  \\
        & $z_{\rm DES}$ & 19.17$\pm$0.01 & 20.87$\pm$0.02 & -- \\
        & $Y_{\rm DES}$ & 19.06$\pm$0.03 & 20.84$\pm$0.10 & -- \\
VHS$^a$ & $J$ & 18.87$\pm$0.05 & 20.37$\pm$0.08 & --  \\
        & $K_{\rm s}$ & 18.67$\pm$0.05 & 20.25$\pm$0.10 & -- \\
UKIDSS$^b$ & $J$ & -- & -- & 20.88$\pm$0.21 \\
catWISE & $W1$ & 18.37$\pm$0.02 & 20.46$\pm$0.07 & 20.40$\pm$0.10 \\
        & $W2$ & 18.36$\pm$0.03 & 20.49$\pm$0.13 & 20.87$\pm$0.29 \\
allWISE$^c$ & $W3$ & 17.39$\pm$0.27 & -- & -- \\
SEIP$^e$& 3.4\textmu m & -- & -- & 20.96$\pm$0.07 \\
        & 4.6\textmu m & -- & -- & 20.68$\pm$0.08 \\

    \hline
    \hline

    \end{tabular}
    \tablefoot{All the values reported have been corrected for Galactic extinction.
    ($^a$) VISTA Hemisphere Survey (VHS; \citealt{McMahon2021});
    ($^b$) UKIRT Infrared Deep Sky Survey (UKIDSS; \citealt{Lawrence2007});
    ($^c$) all-sky WISE (allWISE; \citealt{Cutri2014};)
    ($^d$) Spitzer enhanced imaging products (SEIP; \citealt{Spitzer2021}).}
    \label{tab:photom_data}
\end{table*}

\subsection{Spectroscopic identification}
In order to confirm the high-redshift nature of these three candidates, we observed them with dedicated spectroscopic observations in the range $\sim6000-9000$~\AA, using different telescopes, given their wide range of position in the sky as well as of optical magnitudes. In Table \ref{tab:spec_obs} we summarise the main parameters of the observations.

\begin{table*}
    \caption{Description of the spectroscopic observations obtained for the three new QSOs presented in this work.}
    \centering
       \addtolength{\tabcolsep}{-0.3em}
    \begin{tabular}{lccccccccc} 
        Name & R.A. & Dec. & $z$\_mag & Telescope & Instrument & Grating &  Exp. time &  Date (2023)& \\ 
        \hline
        \hline
        PSO~J0202$-$17  & 30.618886  & $-$17.141042 & 19.22 & AAT &  KOALA & 385R    & 3960 s &  Aug 21 & \\
        \hline
        DES~J0209$-$56 & 32.320529  & $-$56.447344 & 20.87 & VLT &  FORS2 & 600I+25 & 4680 s & Sept 18 + Sept 20 & \\ 
        \hline
        {PSO~J1011$-$01}  &{152.981424} & {$-$1.514636}  & {21.19}  & Gemini-S &  GMOS & R400 & 4800 s  & May 20 + June 14 + June 16 & \\ 
        & &  & &  LBT &  LUCI & {\it zJspec} & 7590 s  & Feb 17 (2024) & \\ 

        \hline
        \hline
    \end{tabular}
    \tablefoot{The magnitudes in the $z$-band are from DES in the case of DES~J0209$-$56 and from Pan-STARRS in the case of PSO~J0202$-$17 and PSO~J0202$-$17.} 
    \label{tab:spec_obs}
\end{table*}

PSO~J0202$-$17 was observed with the Anglo Australian Telescope (AAT; ID O23B005, P.I. Ighina) on August 21, 2023, for a total of 3960~sec. Observations were carried out with the 2df+AAOmega Spectrograph and the Kilofibre Optical AAT Lenslet Array (KOALA) using the 385R grating. In order to reduce the impact of dead pixels, we divided the observations in three dithered exposures, where the offset was given by an integer number of lensets. The raw spectra were reduced using the 2dfdr data reduction software following the KOALA manual\footnote{Available at: \url{https://aat.anu.edu.au/science/instruments/current/koala/manual}.}. 

DES~J0209$-$56 was observed with the Very Large Telescope (VLT; ID 111.24Q5, PI Ighina) for a total of 4680~sec, half on the September 18 and half on September 20, 2023. We used the FORS2 instrument with the 600I+25 grating and performed a standard data reduction with the ESOreflex software \citep{Freudling2013}.

PSO~J1011$-$01 was observed with the Gemini-South telescope (Gemini-S; ID GS-2023-DD-108, P.I. Ighina) for a total of 4800~sec divided across May 20 and June 14 and 16, 2023. The overall time was divided in 12 exposures, four with the central wavelength of the grism at 7900~\AA, four at 8000~\AA \, and four at 8100\AA, in order to cover the spectral gap in the detector as well a faulty part of the CCD at the time of observing. 
Data reduction was performed using the IRAF Gemini package and following the instructions reported in the GMOS Data Reduction Cookbook (Version 1.1; Tucson, AZ: National Optical Astronomy Observatory; Shaw, R. A. 2016)\footnote{Available at: \url{http://ast.noao.edu/sites/default/files/GMOS\_Cookbook}.}.

\begin{figure}
        \includegraphics[width=\hsize]{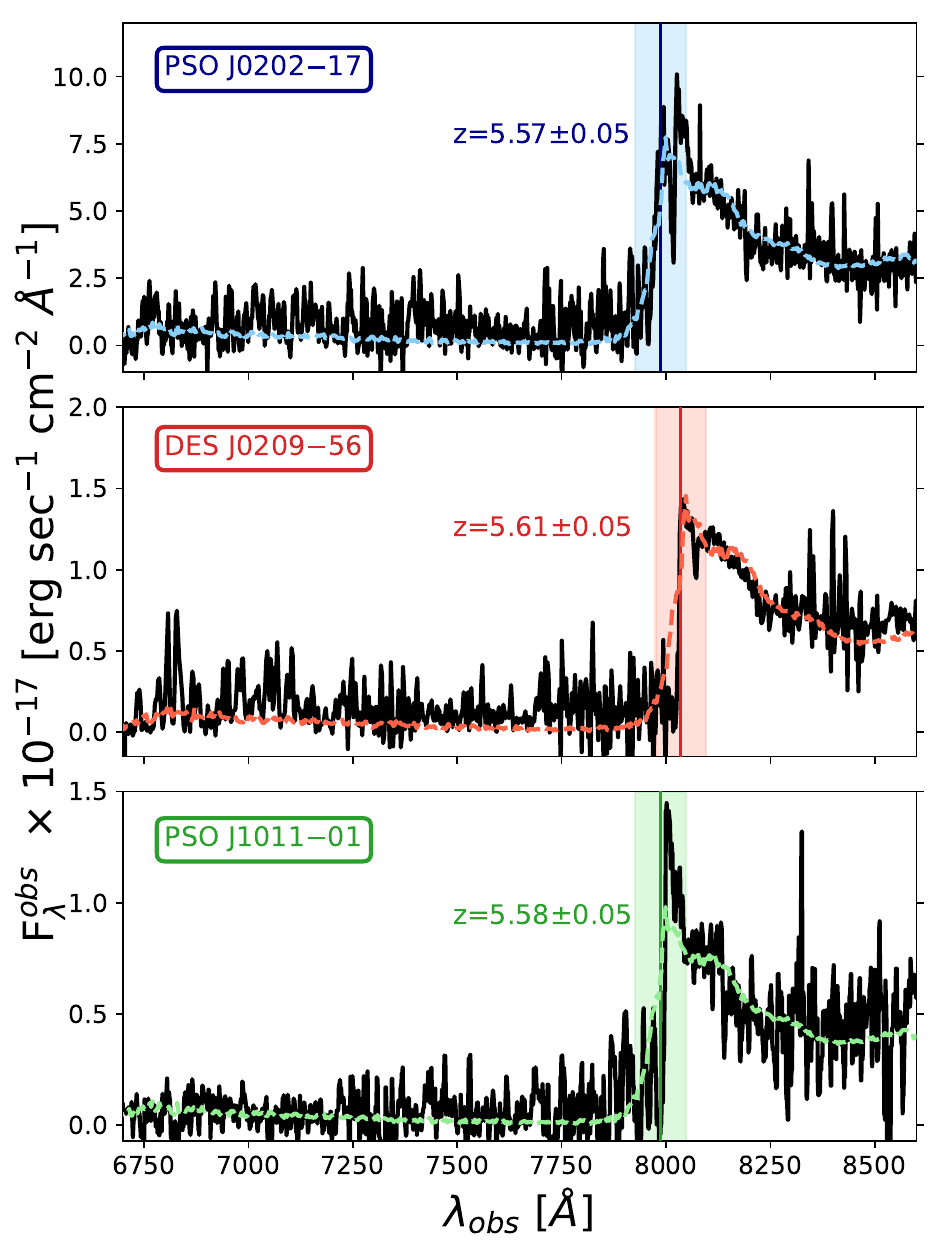}
    \caption{Optical/NIR discovery spectra around the  observed Ly$\alpha$ emission line of the new sources presented in this work. Spectra have been normalised to the corresponding $z$-band magnitude corrected for Galactic absorption. The significant drop at wavelengths shorter than the Ly$\alpha$ confirms the high-redshift nature of all the three objects. The dashed line in each panel shows the composite spectrum derived by {\protect \cite{Banados2016}} at the best-fit redshift of each source. The shaded regions show a variation of the Ly$\alpha$ wavelength corresponding to $\sigma_{z}\pm$0.05.}
    \label{fig:opt_spec}
\end{figure}

We report in Fig. \ref{fig:opt_spec} the final optical spectra obtained for the three targets normalised to the available magnitudes in the $z$-band (corrected for Galactic absorption). In all cases, the large drop in flux observed around $\sim$8000\AA \, confirms the high-$z$ nature of these QSOs via evidence for a Lyman break. 
In order to have an estimate of their redshift we performed a fit using a high-$z$ QSO template around the Ly$\alpha$ break (7600--8600~\AA\, observed range), since it is the only prominent feature observed in all the spectra. 
In particular, we considered the three types of composite spectra derived by \cite{Banados2016} from $5.5<z<6.5$ QSOs (strong Ly$\alpha$, weak Ly$\alpha$ and all the sources). For all the three objects, the spectrum obtained from the combination of all the high-$z$ QSOs found in \cite{Banados2016} resulted in a better fit. In this way we found: $z=5.57$ for PSO~J0202$-$17, $z=5.61$ for DES~J0209$-$56 and $z=5.58$ for PSO~J1011$-$01. As uncertainty on these estimates, we consider $3\times20$~\AA \, ($\sigma_{z}=\pm0.05$), which roughly corresponds to the width of the observed drops (i.e. from the Ly$\alpha$ line to a zero flux; see Fig. \ref{fig:opt_spec}). 

In the specific case of PSO~J1011$-$01, we also obtained dedicated observations in the NIR with the LBT Utility Camera in the Infrared (LUCI; IT-2023B-022; P.I. Ighina) in order to uncover the CIV broad-emission line at 1549.5~\AA \, in the rest frame ($\sim10100$~\AA \, in the observed frame), which, as described below, can be used in to constrain the properties of the central SMBH. Observations were performed in binocular mode on February 17, 2024, for a total of 7590~s (divided into 46 exposure of 165~sec each) using the {\it zJspec} grism (i.e. covering the wavelength range $\sim0.96-1.3$~\textmu m). Data reduction was performed using the SIPGI pipeline\footnote{The corresponding manual can be found at the following link: \url{https://pandora.lambrate.inaf.it/docs/sipgi/}.} \citep{Gargiulo2022}. In particular, we observed the HD82476 star in order to correct for telluric absorption and to flux calibrate the final spectrum. The wavelength calibration was performed using the sky emission lines, reaching an RMS~$\sim$~0.23~\AA \, for both LUCI1 and LUCI2.

\begin{figure}
        \includegraphics[width=\hsize]{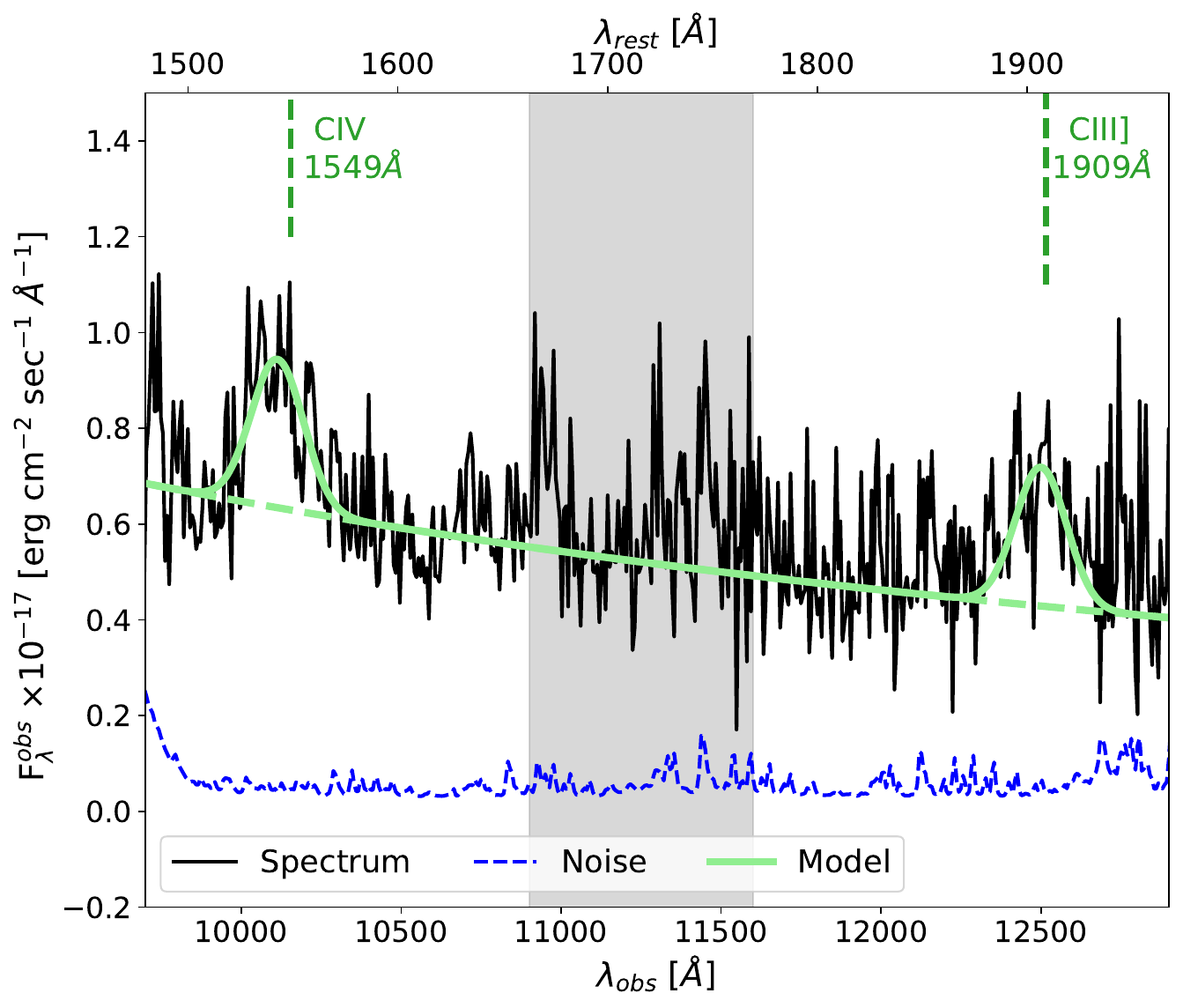}
    \caption{Combined LUCI1+LUCI2 spectrum of PSO~J1011$-$01 around the CIV and CIII] emission lines. The spectrum (black solid line) is binned at 6.5\AA, that is, three times the native resolution. The dashed blue line shows the noise of the final spectrum derived from the data reduction. The solid light green line is the best-fit model of the continuum plus the CIV and CIII] emission lines. The shaded grey area indicates wavelengths heavily affected by telluric absorption that were not considered during the fit.}
    \label{fig:CIV_line}
\end{figure}

In the final spectrum, Fig. \ref{fig:CIV_line}, two emission lines are visible, CIV1549.5\AA \, and CIII]1908.7\AA, which we used in order to have an independent estimate of the redshift of the PSO~J1011$-$01 system. In order to model the continuum plus the emission lines, we considered a simple power law with two Gaussian functions, one for each line. Even though both these lines can present complex shapes (due to, e.g., outflows and/or blending with other lines; \citealt{Denney2016, Marziani2022}), given the limited S/N of the spectrum and the scope of this work, we limited their analysis to a simple, yet satisfactory, model. During the fit we considered the entire  $\sim9600-13000$~\AA \, range, with the exception of the $\sim10900-11600$~\AA \, spectral range since it was heavily affected by telluric adsorptions. From the central position of the CIV and CIII] emission lines, we estimated the redshift of the system to be $z=5.54\pm0.03$ and $z=5.55\pm0.04$, respectively. In both cases the best-fit redshifts obtained are smaller compared to the one derived above from the fit to the continuum and drop around the Ly$\alpha$ line. This difference could be due to the presence of a blue-shifted component, especially in the CIV emission line. Indeed, this line often presents a systematic shift, with respect to other emission lines, towards bluer wavelengths (see, e.g., \citealt{Sun2018,Shen2019}) and the strength of this shift has been found to anti-correlate with the intensity of the line (e.g. \citealt{Ge2019}) and to correlate with the intensity of the continuum emission (e.g. \citealt{Shen2016}). For this reason, we also considered the typical blue-shift observed in $z\gtrsim6$ QSOs, $\sim2200$~km~sec$^{-1}$  (or $\Delta z\sim0.05$; median value in the sample analysed by \citealt{Mazzucchelli2023}), and added it in quadrature to the statistical uncertainty of redshift derived from the CIV line, obtaining $z=5.54\pm0.06$. 
Since all the three redshift measurements available for PSO~J1011$-$01 can potentially be affected by a systematic effect that is hard to quantify with the data currently in hand (absorption from the IGM for the Ly$\alpha$, blue-shift for the CIV and blending for the CIII] line), we considered the weighted average of these values as best-fit estimate for the redshift of this object. The final redshift that we obtained for the PSO~J1011$-$01 QSO and that we use throughout the rest of the paper is therefore $z=5.56\pm0.03$. While the exact redshift value does not impact the discussion presented in this work, further observations in the sub-millimetre band (targeting, e.g., CO or [CII]) will provide a more accurate estimate.

In Sect. \ref{sec:mass_BH} we also use the CIV emission line in order to constrain the properties of the BH hosted in this system. The best-fit parameters obtained for the CIV line are reported in Table \ref{tab:CIV_line}. Following \cite{Diana2022}, uncertainties were computed through Monte Carlo simulations with a Gaussian-distributed noise given by the RMS of the observed spectrum. We considered the standard deviation of the best-fit distribution parameters derived from all the simulated spectra as the final uncertainty on that given parameter.

\begin{table*}
 \caption{Results of the fit to the CIV broad-emission lines in PSO~J1011$-$01. This includes the FWHM, the line luminosity, and the rest-frame equivalent width, indicated as REW.}
        \centering
\begin{tabular}{cccccccccc}
\hline
\hline
FWHM  & L$_{\rm 1350\AA}$ & L$_{\rm line}$ & REW & M$_{\rm BH}$ & $\lambda_{\rm Edd}$\\
(km sec$^{-1}$)  &  erg sec$^{-1}$ & erg sec$^{-1}$ & \AA & 10$^9$~M$_\odot$ \\
\hline
6140$\pm$590 & 1.4$\pm$0.2$\times10^{46}$ & 3.1$\pm$0.3$\times10^{44}$ & 15.5$\pm1.3$ & 2.2$\pm$0.5 & 0.16$\pm$0.07\\
\hline
\hline
 \end{tabular}

 \tablefoot{The continuum emission was estimated from the photometric measurements available for PSO~J1011$-$01. The uncertainties of the mass and accretion rate reported in this table do not take the scatter of the scaling relations used for their derivation into account.}
\label{tab:CIV_line}

\end{table*}

\section{Radio and X-ray observations}

In order to constrain the multi-wavelength spectral energy distribution (SED) of these newly identified QSOs we performed dedicated observations in addition to all the radio observations already available as part of public surveys. In particular, we observed all three targets with the Australia Telescope Compact Array (ATCA), in the radio band, and with the \textit{Swift}-XRT telescope, in the X-ray band. 

\subsection{Radio data}
In this subsection we describe the radio data available for the three sources discussed in the paper. In the first subsection we list the radio surveys publicly available where at least one of the QSOs discussed in this paper is detected. In the second subsection we present new radio data for all the three QSOs obtained with dedicated ATCA observations. 
All the three sources have multiple radio observations from publicly available surveys, typically in the frequency range $\sim$0.1-3~GHz as well as dedicated ATCA observations up to 9~GHz, resulting in a radio coverage over almost two decades in frequency.

\subsubsection{Radio surveys}
 In Table \ref{tab:radio_data} all the flux density estimates from public radio surveys are reported for each source. In particular, these values come from the TIFR GMRT Sky Survey (TGSS; \citealt{Intema2017}), the Sydney University Molonglo Sky Survey  (SUMSS; \citealt{Mauch2003}), the RACS-low survey \citep{Hale2021}, the RACS-mid survey \citep{Duchesne2023,Duchesne2024}, the NRAO VLA Sky Survey (NVSS; \citealt{condon1998}), the Faint Images of the Radio Sky at Twenty-Centimeters (FIRST; \citealt{Becker1995}) and the VLASS (\citealt{Lacy2020}). In the case of DES~J0209$-$56 we also include the measurement from the Evolutionary Map of the Universe (EMU) based on the source-list catalogue reported on CASDA, where we also added a 5\% uncertainty in quadrature to the final error. We note that for the VLASS measurements, we performed a fit on the quick look images of the different epochs available using the CASA software. Following the recommendations reported in \cite{Gordon2020}, we corrected the integrated flux densities by adding in quadrature 10\% for the epoch 1.1 and 3\% for the other epochs. Furthermore, in all the estimates from the different epochs, we always consider an additional 10\% in the uncertainty added in quadrature to the statistical error from the fit. Moreover, both PSO~J0202$-$17 and DES~J0209$-$56 have been detected n the recent 2$^{\rm nd}$ data release of the Extra-Galactic All-Sky Murchison Widefield Array (MWA) Extended Survey (GLEAM-X; \citealt{Ross2024}). However, in the case of DES~J0209$-$56 the S/N was not enough to perform a good fit to its radio emission in each of the twenty 7~MHz channels imaged with the MWA. For this reason, as measurement of the low-frequency emission of DES~J0209$-$56 we considered its peak flux density reported in the five wider bands (30~MHz each) reported in the GLEAM-X catalogue \citep{Ross2024} with the corresponding off-source RMS as uncertainty. From the fit of these simultaneous low-frequency data only.
 In Table \ref{tab:gleamX} we report the simultaneous measurements from the GLEAM-X survey in all the twenty 7~MHz bands for PSO~J0202$-$17 and in the five wide 30~MHz bands for DES~J0209$-$56.


\begin{table*}
    \caption{Flux density values of the three sources discussed in this work as measured in different public surveys and dedicated ATCA observations.}
    
    \centering
    \begin{tabular}{lcccc}
    Survey/ & Frequency &  \multicolumn{3}{c}{Integrated flux density}\\
    Telescope & (GHz) & \multicolumn{3}{c}{(mJy)}\\
    \hline\hline
         &  &   PSO~J0202$-$17 & DES~J0209$-$56 & PSO~1011$-$01 \\
         \hline
        TGSS & 0.150 & 42.6$\pm$8.1 & -- & 7.4$\pm$2.1$^*$\\
        SUMSS & 0.843 & -- & 15.4$\pm$0.9 & -- \\
        RACS-low & 0.888 & 19.8$\pm$2.0 & 18.2$\pm$1.8 & 8.5$\pm$1.3 \\
        EMU & 0.944 & -- & 18.9$\pm$0.9 & -- \\
        RACS-mid & 1.37 & 13.6$\pm$0.9 & 17.8$\pm$1.1 & 7.7$\pm$0.6 \\
        NVSS & 1.4 & 12.5$\pm$0.6 & -- & 6.0$\pm$0.5\\
        FIRST & 1.4 & -- & -- & 6.7$\pm$0.3\\
        VLASS--1 & 3 & 14.3$\pm$1.5 & -- & 6.1$\pm$0.7\\
        VLASS--2 & 3 & 14.9$\pm$1.6 & -- & 5.8$\pm$0.7\\
        VLASS--3 & 3 & -- & -- & 5.2$\pm$0.6$^1$\\
        \hline
         ATCA & 2.1 & 16.1$\pm$0.8 & 13.0$\pm$0.7 & -- \\
              & 5.5 & 19.1$\pm$0.9 & 10.6$\pm$0.5 & 4.9$\pm$0.5$^\dagger$ \\
              & 9 & 13.8$\pm$0.7 & 9.5$\pm$0.5 & 2.7$\pm$0.3$^\dagger$\\
    \hline\hline

    \end{tabular}
    \tablefoot{The `--' symbol indicates that the specific source is not currently covered by the given survey. 
    $^*$ we considered the peak surface brightness estimated from the TGSS image of PSO~J1011$-$01; 
    $^\dagger$ derived by fitting the visibilities assuming a point source model.}
    \label{tab:radio_data}
\end{table*}

\begin{table*}
    \caption{Peak flux density of PSO~J0209$-$17 and DES~J0209$-$56 as reported in the GLEAM-X catalogue {\protect\cite{Ross2024}}.}
    \centering
    \begin{tabular}{cccccccc}
\multicolumn{4}{c}{PSO~J0202$-$17} &&& \multicolumn{2}{c}{DES~J0209$-$56} \\
        Frequency & Peak flux density  & Frequency & Peak flux density &&& Frequency & Peak flux density \\
         (MHz) & (mJy~beam$^{-1}$) & (MHz) & (mJy~beam$^{-1}$) &&& (MHz) & (mJy~beam$^{-1}$)\\

         \hline
76  &    83  $\pm$  7 & 158 &    48  $\pm$  2   &&& 87  & 20 $\pm$  5  \\
84  &    89  $\pm$  7 & 166 &    42  $\pm$  2   &&& 118 & 16 $\pm$  3  \\
92  &    55  $\pm$  6 & 174 &    46  $\pm$  2   &&& 154 & 15 $\pm$  2  \\
99  &    66  $\pm$  5 & 181 &    44  $\pm$  2   &&& 185 & 16 $\pm$  2  \\
107 &    61  $\pm$  4 & 189 &    46  $\pm$  2   &&& 215 & 14 $\pm$  2  \\
115 &    64  $\pm$  4 & 197 &    40  $\pm$  2   \\
122 &    54  $\pm$  4 & 204 &    40  $\pm$  2   \\
130 &    41  $\pm$  4 & 212 &    35  $\pm$  2   \\
143 &    50  $\pm$  3 & 220 &    43  $\pm$  2   \\
151 &    47  $\pm$  3 & 227 &    35  $\pm$  2   \\
        \hline
        \hline

    \end{tabular}
 \tablefoot{Estimates for PSO~J0209$-$17 are reported in each of the twenty 7~MHz wide bands, while for DES~J0209$-$56 we only report the values in the five 30~MHz wide bands (due to its lower S/N). The uncertainties on the DES~J0209$-$56 flux densities correspond to the off-source RMS in the given band.}
    \label{tab:gleamX}
\end{table*}

\subsubsection{ATCA observations and analysis}
\begin{figure}
        \includegraphics[width=\hsize]{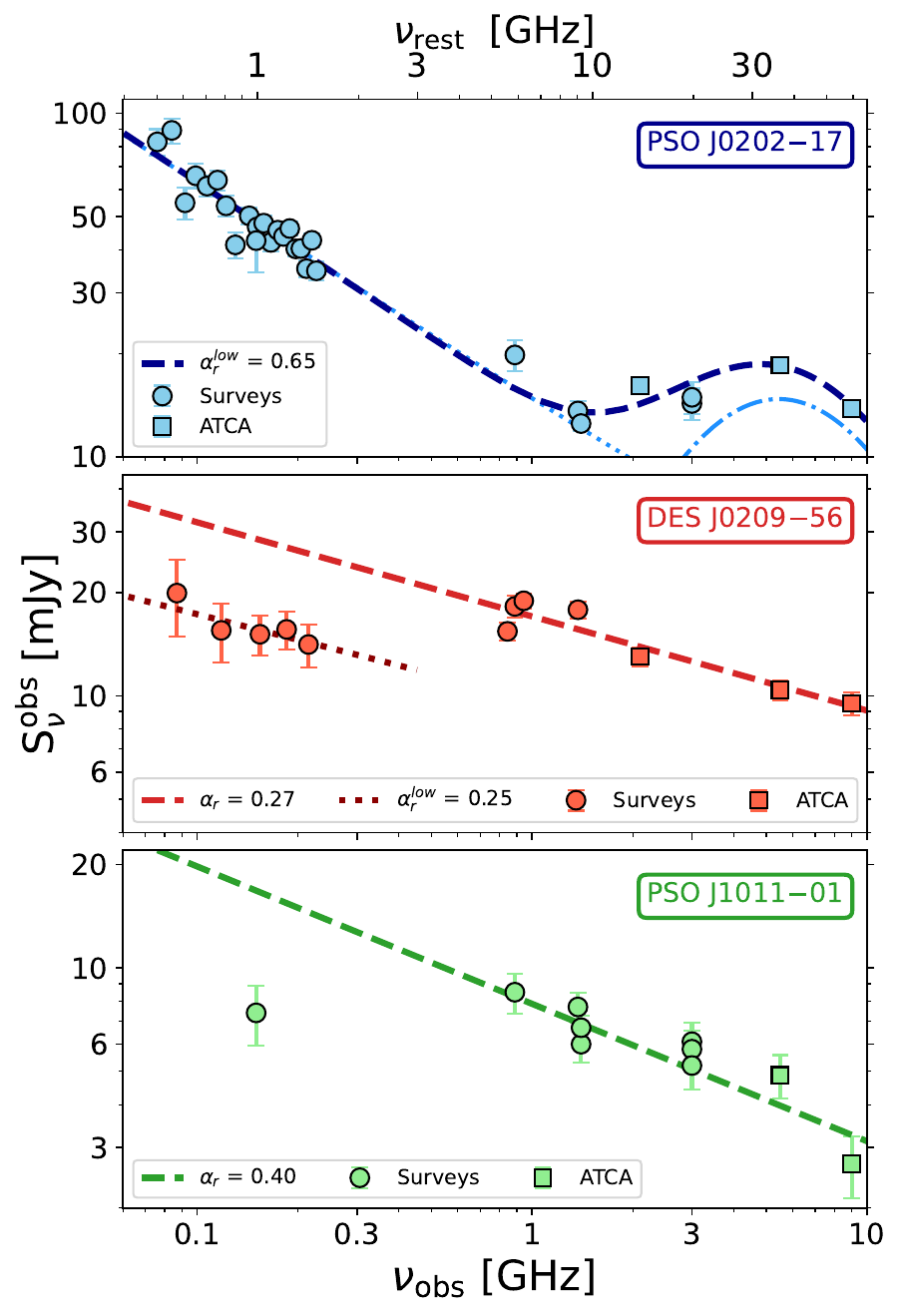}
    \caption{Radio spectra over the $\sim0.1-10$~GHz observed frequency range of the three sources described in this work. The rest-frame frequencies (top labels) were computed assuming a common redshift, $z=5.6$, for all the three objects. The majority of the data points are from public surveys (circles), while the 2.1, 5.5 and 9~GHz measurements are from dedicated, simultaneous observations with ATCA (squares). The dashed line in each panel indicates the best-fit model obtained for the different objects as described in the text. The dotted line in the central panel corresponds to the best-fit power law obtained from the GLEAM-X data only, while the dotted and the dashed-dotted lines in the top panel show the two spectral components used during the fit, as described in the text.}
    \label{fig:radio_spec}
\end{figure}

Dedicated ATCA observations were performed on June 23, 24, and 25, 2023, with the 6D configuration (the most extended; longest baseline $\sim$6000~m) under the project C3535 (P.I. Ighina). Observations were carried out at 5.5 and 9~GHz for all targets and at 2.1~GHz for PSO~J0202$-$17 and DES~J0209$-$56. Observing segments at different frequencies were spread over three runs within 48~h, in order to have a well sampled UV coverage. Given the short time scale ($\sim$8~h in the rest frame) as well as the consistency between the flux densities from each run, we consider the 2.1, 5.5 and 9 GHz flux densities as being  simultaneous during the rest of the paper. To process the data (calibration and imaging), we used the \texttt{MIRIAD} data-reduction package \citep{Sault1995} following a standard reduction. In particular, for imaging, we adopted a robust parameter of 0.5. clean the images we made use of the \texttt{mfclean} task and used the model derived from cleaning in order to self-calibrate the data (two iterations, only on the phases). Finally, we performed a 2D Gaussian fit on the target using the Common Astronomy Software Applications package \citep[\texttt{CASA};][]{Mcmullin2007}. Both PSO~J0202$-$17 and DES~J0209$-$56 are unresolved according to the fit. In the case of PSO~J1011$-$01, we did not perform imaging, since the expected beam for an equatorial source with an east-west array like ATCA would be too elongated. Therefore, in order to estimate the 5.5 and 9~GHz flux densities of this source, we performed a point-source fit in the UV plane (using the task \texttt{UVFIT} in \texttt{MIRIAD}). We note that no other radio sources are visible at 3~GHz in the VLASS survey within the ATCA primary beam ($\sim$5-10$'$, depending on the frequency) and that the source appears point-like in all the other observations.
Moreover, in order to ensure that this is a reliable method, we performed the same fit to the other two targets, finding an agreement between the best-fit values from the 2D Gaussian fit in the image plane and the fit in the UV plane within $\sim$3\% of the total flux. In the following, we assume a conservative 10\% for the uncertainty of the ATCA measurements of PSO~J1011$-$01. For all the sources, we consider a further 5\% error, added in quadrature, to account for uncertainties related to the absolute flux scale and the reduction process. 
These measurements are reported in Fig. \ref{fig:radio_spec} and in Table \ref{tab:radio_data} together with the ones available from public radio surveys.

\subsection{Swift-XRT observations and analysis}
Dedicated \textit{Swift}-XRT observations were performed on PSO~J0202$-$17, DES~J0209$-$56 and PSO~J1011$-$01 (P.I. Ighina) in multiple segments between March and December 2023 for a total time for all the three targets of 114~ksec (see Table \ref{tab:x-ray_fit}). Individual segments were processed and combined with the specific \textit{Swift} software included in the package \texttt{HEASOFT} (v. 6.32; \citealt{Evans2009}) available from the public archive\footnote{\url{https://www.swift.ac.uk/user_objects/}}. PSO~J0202$-$17 and DES~J0209$-$56 were both detected with 42 and 19 net photons in the 0.3--10~keV energy band (in a circular region of radius 20~pixels or $\sim45''$) over a background of 12.5 and 9.4 photons, respectively. In both cases, the radio and optical positions are consistent with the standard, PSF-fitted \textit{Swift}-XRT positions (uncertainty $\sim4-6''$, at 90\% confidence level). However, in the case of PSO~J1011$-$01, there is not a significant detection. In particular, the number of photons in a region of 10$''$ around the optical position of the QSO  and in the 0.3--10~keV range and is four, which is consistent at the 90\% confidence level with the expected counts from the background ($\sim$4.3).

\begin{figure}
{\centering
        \includegraphics[width=0.9\hsize]{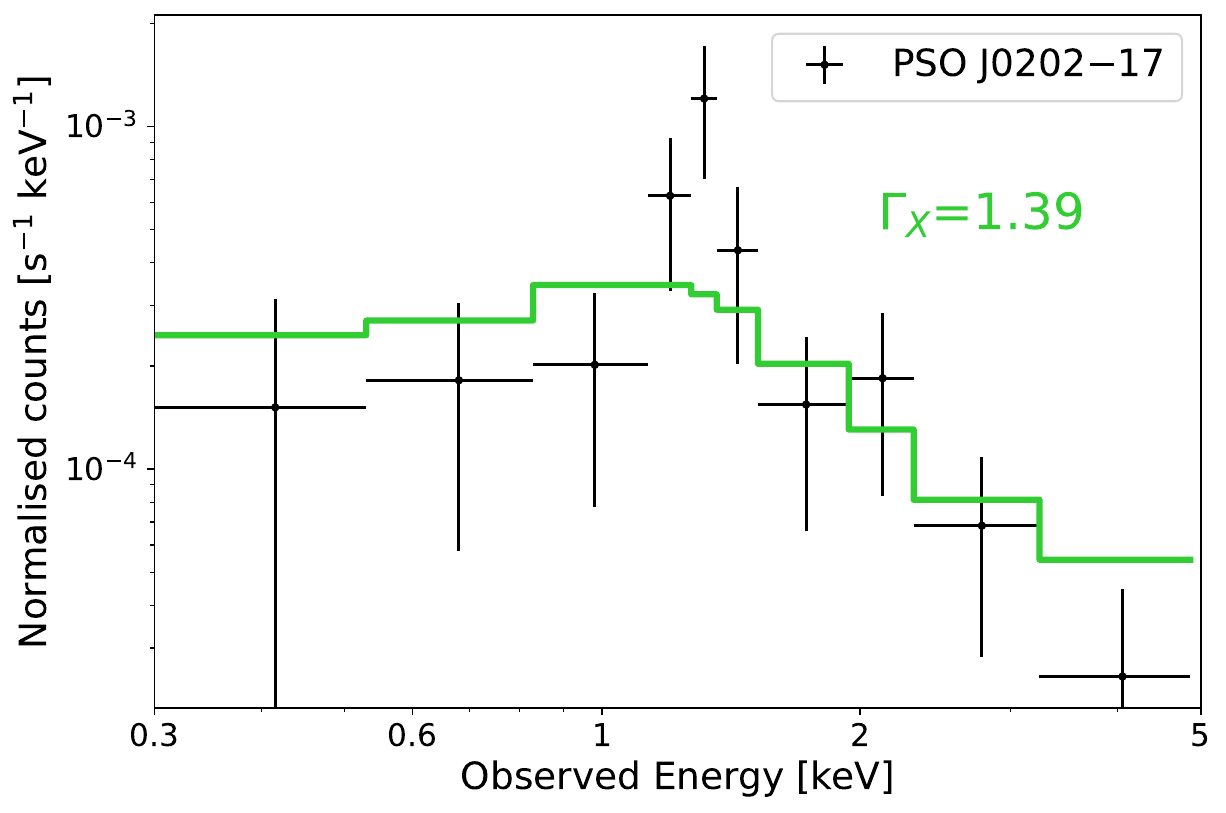}
        \includegraphics[width=0.9\hsize]{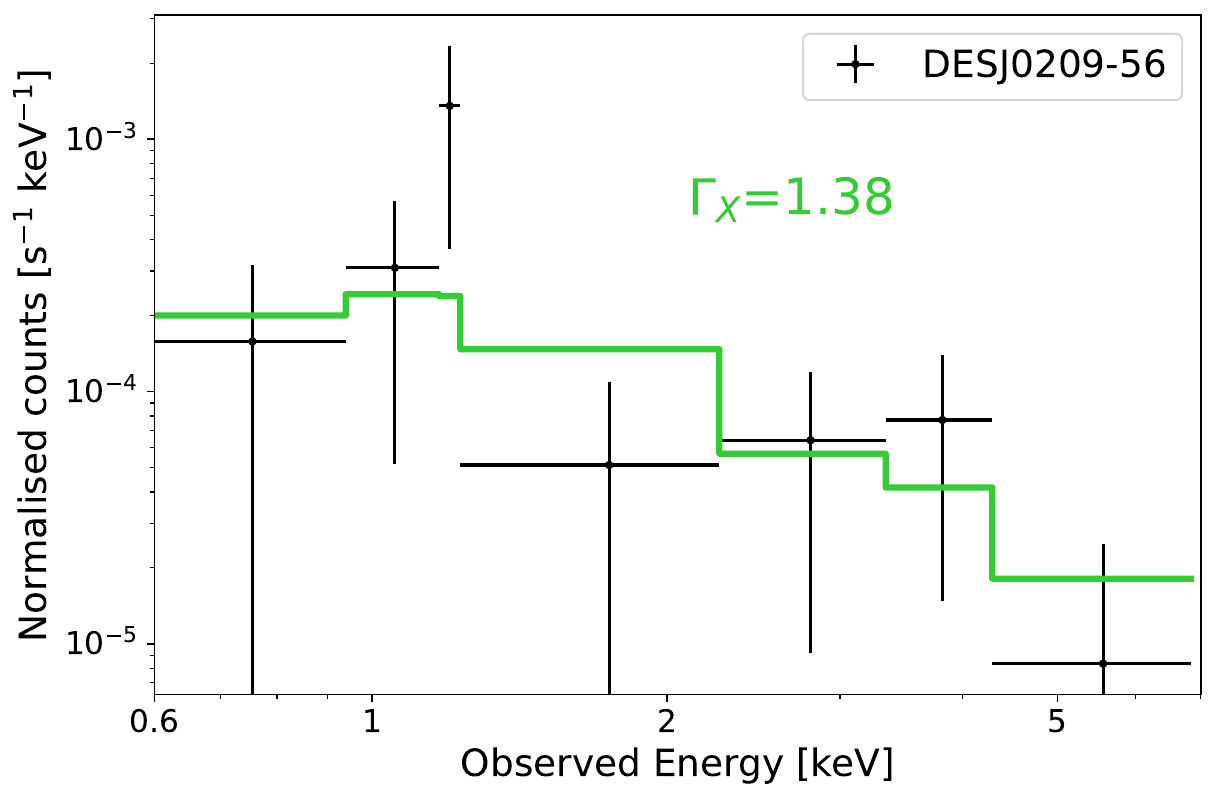}
    \caption{X-ray spectra of PSO~J0202$-$17 and DES~J0209$-$56 obtained with the \textit{Swift}-XRT telescope. The solid green line shows the best-fit power law with Galactic absorption model. }
    \label{fig:x_spec}
    }
\end{figure}

\begin{figure}
    
{\centering
        \includegraphics[width=0.9\hsize]{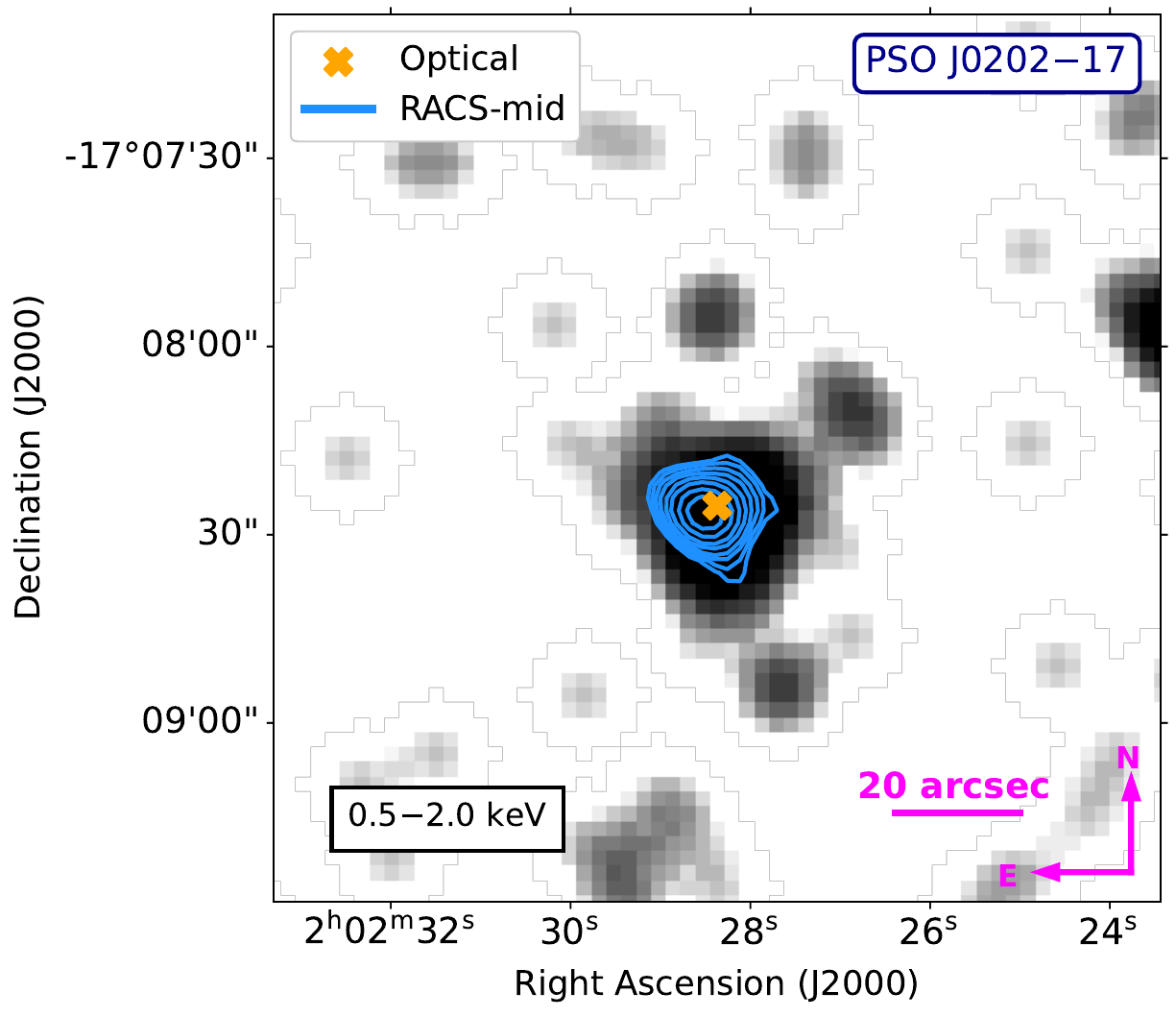}\\
        \includegraphics[width=0.9\hsize]{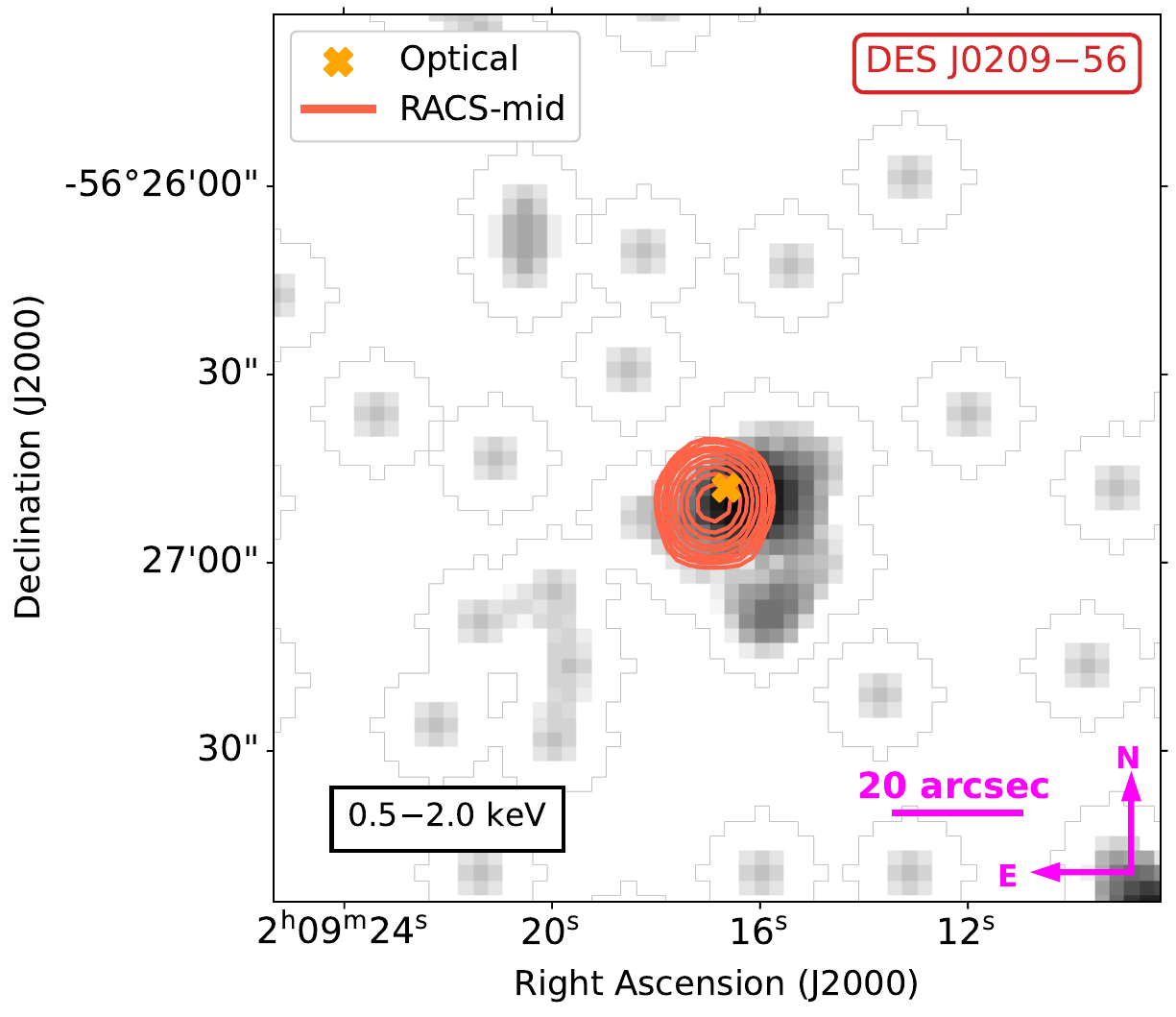}\\
        \includegraphics[width=0.9\hsize]{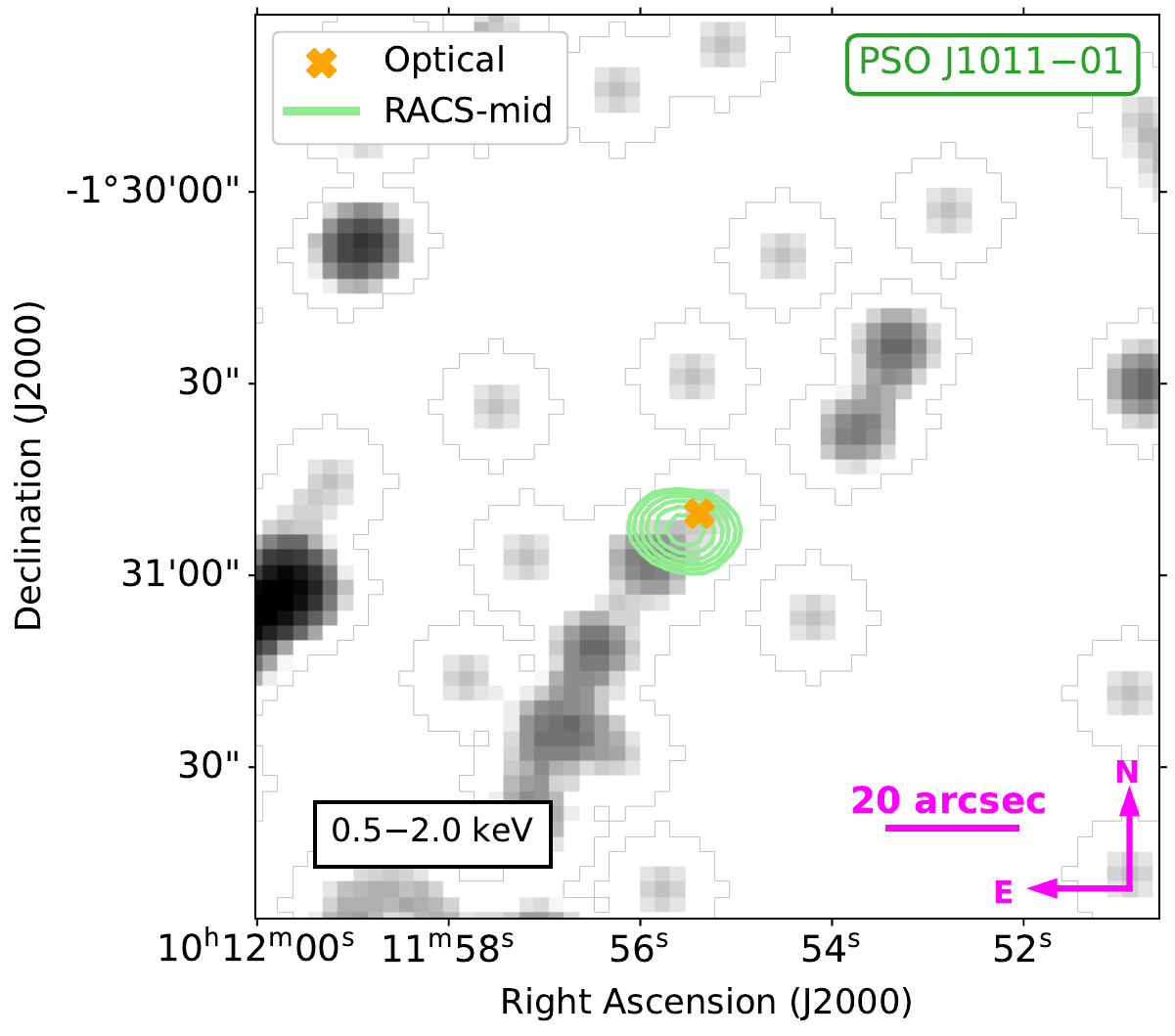}
    \caption{X-ray images in the 0.5--2~keV energy range of the three high-$z$ sources described in this paper obtained with the \textit{Swift}-XRT telescope and smoothed with a Gaussian function with $\sigma=2$~pixels ($\sim4.7''$). The orange cross indicates the optical position of the QSO, while the contours are from the RACS-mid images and are drawn starting from $\pm5~\times$~off-source RMS and spaced by factors of $\sqrt{2}$.}
    \label{fig:x_images}
    }
\end{figure}

The data analysis for the spectra of PSO~J0202$-$17 and DES~J0209$-$56 was performed with the \texttt{XSPEC} software (v12.11.1; \citealt{Arnaud1996}). In particular, we adopted a simple power-law model with Galactic absorption only (N$_{\rm H}$ values from \citealt{HI4PI2016}). To perform the fit, we adopted the c-statistics \citep{Cash1979}, which is more suited for low-number datasets, such as ours, since it does not require binning of the data. We report in Table \ref{tab:x-ray_fit} the results of the fit and in Fig. \ref{fig:x_spec} the spectra obtained from the analysis with their best-fit models. In Fig. \ref{fig:x_images} we present the X-ray images, retrieved from the \textit{Swift}-XRT archive, of all the three targets together with the radio contours from the RACS-mid images overlaid.
    
\begin{table}
        \centering
        \caption{Best-fit values obtained from the fit of the \textit{Swift}-XRT X-ray spectra of the targets discussed in this work.} 
        \label{tab:x-ray_fit}

\begin{threeparttable}
\centering
 \addtolength{\tabcolsep}{-0.4em}
\begin{tabular}{ccccccc}
\hline
\hline
$\Gamma_{\rm X}$    &   Flux$^\mathrm{a}$ &   Lum.$^\mathrm{b}$  & photon & exp. time & cstat/d.o.f.\\
   &   0.5--10 keV &   2--10 keV  & counts & ksec & \\
\hline
&& \multicolumn{2}{c}{PSO~J0202$-$17} & & \\
1.39$^{+0.48}_{-0.50}$  &       5.2$_{-1.8}^{+1.7}$  &   3.4$_{-0.5}^{+1.8}$   & 42 & 54 & 31 / 38\\
\hline
&& \multicolumn{2}{c}{DES~J0209$-$56} & & \\
1.38$^{+0.77}_{-0.78}$  &       3.4$_{-1.4}^{+1.0}$  &   2.6$_{-0.7}^{+1.6}$   & 19 & 30 & 19 / 19\\
\hline
&& \multicolumn{2}{c}{PSO~J1011$-$01} & & \\
--  &   $<0.8^c$ &   $<1.1^c$   & $<3.2$ & 30 & -- \\

\hline
\hline
\end{tabular}

 \tablefoot{The fit was performed assuming a simple power law with Galactic absorption from {\protect\cite{HI4PI2016}}. Col. (1) best-fit photon index; col. (2) un-absorbed flux in the energy band 0.5--10~keV; col. (3) rest-frame luminosity in the energy range 2--10~keV; col. (4) total number of photons from the target; col. (5) c-statistic and degree of freedom of the fit.
Errors and upper-limits are reported as a 90\% level of confidence.
[a] in units of 10$^{-14}$ erg sec$^{-1}$ cm$^{2}$;
[b] in units of 10$^{45}$ erg s$^{-1}$;
[c] assuming $\Gamma_{\rm X}=1.8$.}

\end{threeparttable}
\end{table}

\section{Notes on individual sources}
\label{sec:ind_obj}

\begin{figure*}
        \includegraphics[width=\hsize]{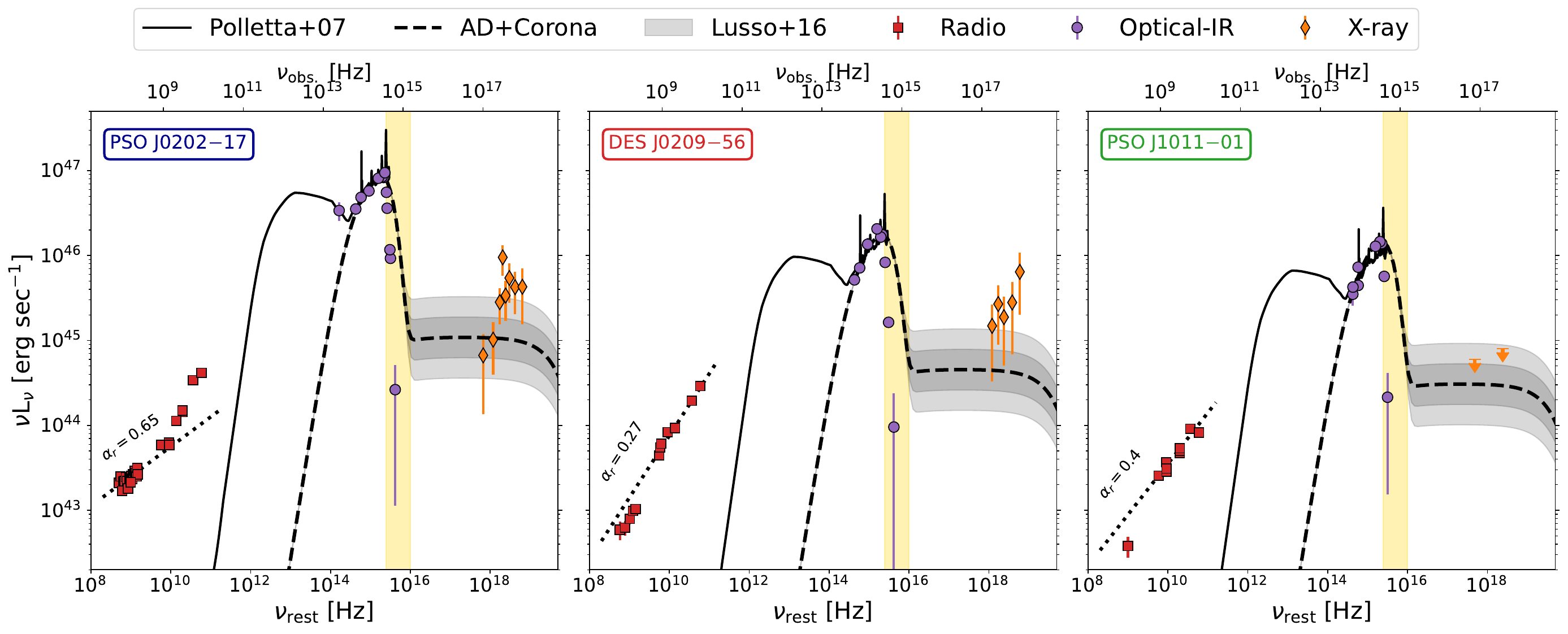}

    \caption{Multi-wavelength rest-frame SED of the three new RL QSOs presented in this work. For each source we show the radio (red squares), optical/NIR (purple circles) and X-ray (orange diamonds) photometry discussed in the text. The black line is a QSO template from {\protect \cite{Polletta2007}}, while the shaded yellow region represents frequencies affected by the hydrogen absorption from the inter-galactic medium (IGM). The dashed line is the best-fit AD model to the optical data points (assuming $\eta\sim0.1$) together with the expected X-ray emission from the optical/UV--X-ray relation found in {\protect \cite{Lusso2016}} (assuming $\Gamma_{\rm X}=2$ and an exponential cut-off at 200~keV). The Corresponding shaded area in the X-rays shows the 1 and 2$\sigma$ scatter of the relation. The dotted black line at low frequency is the best-fit power law to the radio data, as described in the text.}
    \label{fig:SED}
\end{figure*}

In this section we present a brief discussion of the radio and X-ray properties of each source reported in this work. In Fig. \ref{fig:radio_spec} and in Fig. \ref{fig:x_spec} we show their radio and X-ray spectra respectively, while in Fig. \ref{fig:SED} we show their multi-wavelength SED, obtained by the collection of all the multi-wavelength data points available. In this last plot, the dashed black line is given by the sum of the best-fit accretion disc model (with $\eta_{\rm d}=0.1$; see Sect. \ref{sec:mass_BH}) and the expected X-ray emission from the X-ray corona based on the optical/UV-X-ray relation derived in \cite{Lusso2016}, assuming a photon index $\Gamma_{\rm X}=2$ \citep[typical of $z>4$ radio-quiet QSOs; e.g. ][]{Vignali2005,Nanni2017} and an exponential cut-off at 200~keV \citep[e.g.][]{Lanzuisi2019,Bertola2022}. We note that, while the \cite{Lusso2016} relation was derived with a sample of radio-quiet (or non-radio) QSOs, we expect their X-ray emission to be similar to the one in mis-aligned jetted systems, since the X-ray radiation produced by the jets in these systems can be significantly de-boosted (see, e.g., fig. 5 in \citealt{Ighina2019}). Finally, in Table \ref{tab:multi_nu_param} we report the values of different multi-wavelength parameters described below and computed for each object. The fits to the radio data points reported in this section were performed using the \texttt{MrMOOSE} code, described in \cite{Drouart2018a,Drouart2018b}.

\begin{table*}
\caption[Luminosities and spectral indices for three $z\sim5.6$ RL QSOs.]{Multi-wavelength spectral indices and luminosities, radio loudness, and $\tilde{\alpha}_{\rm ox}$ parameter of the three sources discussed in this work.}
\centering

\begin{tabular}{ccccccc}
\hline
\hline

     & & PSO~J0202$-$17  &  & DES~J0209$-$56  &  & PSO~J1011$-$01 \\
\hline 
$\alpha_{\rm r}$ && 0.65$\pm$0.02 && 0.27$\pm$0.02 && 0.40$\pm$0.03\\ 
$\alpha_{\rm o}^{\nu}$ && 0.52$\pm$0.01 && 0.30$\pm$0.04 && 0.27$\pm$0.03\\
${\rm L}_{\rm5GHz}$ [erg~sec$^{-1}$~Hz$^{-1}$] && 1.1$\pm$0.1$\times10^{34}$ && 9.8$\pm$1.0$\times10^{33}$ && 4.6$\pm$0.7$\times10^{33}$\\
${\rm L}_{\rm2500{\normalfont\AA}}$ [erg~sec$^{-1}$~Hz$^{-1}$] && 6.1$\pm$0.3$\times10^{31}$ && 1.1$\pm$0.1$\times10^{31}$ && 8.6$\pm$1.4$\times10^{30}$\\
${\rm L}_{\mathrm{{4400\AA}}}$ [erg~sec$^{-1}$~Hz$^{-1}$] && 7.7$\pm$0.3$\times10^{31}$ && 1.2$\pm$0.1$\times10^{31}$ && 1.2$\pm$0.2$\times10^{31}$\\
${\rm L}_{\rm10keV}$ [erg~sec$^{-1}$~Hz$^{-1}$] && 1.6$\pm$0.1$\times10^{27}$ && 1.1$\pm$0.1$\times10^{27}$ &&  $<3.3\times10^{26}$\\
R && 150$\pm$20 && 830$\pm$90 && 380$\pm$70 \\
$\tilde{\alpha}_{\rm ox}$ && 1.39$\pm$0.02 && 1.22$\pm$0.03 && $>1.33$\\

\hline
\hline
\end{tabular}
\label{tab:multi_nu_param}
\end{table*}

\subsection{PSO~J0202$-$17}
Among the three objects discussed in this work, PSO~J0202$-$17 presents  the brightest emission in all the observed electromagnetic bands and, as a consequence, the different spectral parameters are relatively well constrained. 
In particular, the radio spectrum shows a complex shape (see Fig. \ref{fig:radio_spec}, top).
If we consider the simultaneous observations at low frequencies ($<250$~MHz) from the second data release of the GLEAM-X (\citealt{Ross2024}) survey, the spectrum is consistent with a single power law of decreasing intensity with increasing frequency, whereas, at higher frequencies (2.1, 5.5 and 9~GHz; ATCA data), the emission presents a potential peak between 2.1 and 9~GHz. For this reason, we performed a fit of the radio spectrum of PSO~J0202$-$17 using a double component model given by the sum of a simple power law (describing the emission at low frequency, below $\lesssim1$~GHz) and a curved spectrum (describing the high-frequency emission, above $\gtrsim1$~GHz). Following \cite{Callingham2017}, we adopted the equation

\begin{equation}
\centering
    S_\nu = N_{\rm c} \, \nu^{-\alpha_{\rm c}} \, e^{q\,({\mathrm{ln}} \nu)^2} + N_{\rm low}\, \nu^{-\alpha_{\rm low}} ,
    \label{eq:curved_pl}
\end{equation}where $N_{\rm low}$ and $\alpha_{\rm low}$ are the normalisation and the spectral index of the power-law component, $N_{\rm c}$ is the normalisation of the curved component, $q$ is a measurement of its curvature ($|q|$\textgreater0.2 characterises a significantly curved spectrum; \citealt{Callingham2017}) and $\alpha_{\rm c}$ is the spectral index of the second component without curvature. The peak frequency of this last component is given by $\nu_\mathrm{peak}$ = $e^{\alpha_{\rm c}/2q}$.

In this way, we found a best-fit spectral index at low frequencies of $\alpha_{\rm low}=0.65\pm0.03$ and a high-frequency peak at $\nu_{\rm peak}=5.4\pm2.2$~GHz (with $\alpha_{\rm c}=3.4\pm{0.5}$ and $q=-1.0\pm{0.2}$). These two spectral shapes in the radio spectrum of PSO~J0202$-$17 are likely associated with different emitting regions. Indeed, the low-frequency emission, having a spectral index similar to most of the other RL QSOs (e.g. \citealt{Banados2015,Sotnikova2021}), is probably produced by the more extended regions of the relativistic jets, whereas the high-frequency peaked emission is likely produced in a more compact region that is potentially very young \citep[see, e.g.,][]{Orienti2016}, where the radiation below the peak can be absorbed either by the same emitting electrons or by the surrounding medium \citep[e.g.][]{Odea2021}. We note that PSO~J0202$-$17 is not resolved in any radio image available (the best resolution is from ATCA at 9.0GHz, where the beam size is $\sim1''\times6''$). 
Future VLBI observations of this source will help in understanding the origin, in terms of morphology and distance from the central SMBH, of the two components in the radio spectrum.

In order to derive the best-fit optical/IR spectrum of PSO~J0202$-$17 we considered all the photometric measurements not affected by the IGM available, with the exception of the $W3$ band, since it traces another component (likely the dusty torus) with respect to the accretion disc (see Fig. \ref{fig:SED}). The best-fit spectral index of a power law describing the optical/IR emission is $\alpha_{\rm o}^{\nu}=0.52\pm0.02$. Using the magnitude in the band $W1$ (the closest to $4400$~\AA \, rest frame), the radio spectral index at low frequencies ($\alpha_{\rm r}^{\rm low}=0.65$) and the flux density at 888~MHz (the closest to 5~GHz in the rest frame), we derived the radio-loudness parameter for PSO~J0202$-$17, obtaining ${\rm R}=150\pm$20 (see Table \ref{tab:multi_nu_param}).

PSO~J0202$-$17 is also bright at high energies, in the X-rays, having a rest-frame luminosity L$_{\rm 2-10keV}=3.4_{-0.5}^{+1.8}\times10^{45}$~erg~sec$^{-1}$. In order to quantify the strength of the X-ray emission relative to the optical one in high-$z$ QSOs, we considered the $\tilde{\alpha}_{\rm ox}$ parameter, defined as $\tilde{\alpha}_{\rm ox}$~=~--0.303~log$\frac{L_{10keV}}{L_{2500\text{\normalfont\AA}}}$ \citep{Ighina2019}. In the case of PSO~J0202$-$17 we obtained $\tilde{\alpha}_{\rm ox}=1.39\pm0.02$; where we used $\alpha_{\rm o}^{\nu}=0.52$ and the J-band to compute the luminosity at 2500~\AA \, and the best-fit model of Table \ref{tab:x-ray_fit} for the X-ray luminosity. While the best-fit photon index derived for PSO~J0202$-$17 ($\Gamma_{\rm X} = 1.39^{+0.48}_{-0.50}$; uncertainties at 90\% confidence level) is still consistent with being a blazar (i.e. its relativistic jet is oriented close to our line of sight), according to the classification of \cite{Ighina2019}, the $\tilde{\alpha}_{\rm ox}$ indicates that the observed X-ray emission could still be dominated, or partially produced, by the X-ray corona. 

By looking at the multi-wavelength SED of PSO~J0202$-$17, reported in Fig. \ref{fig:SED}, the observed X-ray emission is still marginally consistent with the one expected from \cite{Lusso2016}. However, given that the observed emission increases as a function of frequency ($\Gamma_{\rm X}<1.9$, at a 90\% confidence level), the radiation at the highest energies sampled by \textit{Swift}-XRT ($\sim$50~keV or $\sim$10$^{19}$~Hz in the rest frame) becomes significantly different from the one expected from the X-ray hot corona. It is therefore likely that in this case the relativistic jet and the hot corona emission have a similar intensity and that we are observing the combination of these two components. Given the overall properties of PSO~J0202$-$17 and the values of the R and $\tilde{\alpha}_{\rm ox}$ parameters in particular, this source is likely an object with relativistic jets slightly mis-aligned, for which relativistic boosting is still present, but not enough to make the X-ray jet emission dominate over the other components (see, e.g., \citealt{Belladitta2023} for a similar object). The VLBI observations will help in constraining important physical parameters, such as the bulk Lorentz factor of the jet \citep[e.g.][]{Spingola2020,Homan2021}, which, together with SED modelling (e.g. \citealt{Ghisellini2015b}), can then be used to have a more reliable classification.

\subsection{DES~J0209$-$56}
Differently from PSO~J0202$-$17, the high-frequency radio spectrum of DES~J0209$-$56 can be described by a single power law (see Fig. \ref{fig:radio_spec}). Indeed, from the fit of the simultaneous ATCA data the derived spectral index is $\alpha_{\rm r}=0.27\pm0.02$. Focusing on the simultaneous, low-frequency ($<250$~MHz)  measurements from the GLEAM-X survey, the best-fit spectral index obtained is $\alpha_{\rm r}^{\rm low}\sim0.25^{+0.16}_{-0.14}$. While this value is consistent with the one measured at higher frequencies, the normalisation is a factor $\sim2$ fainter than what expected from the extrapolation of the ACTA measurements.

In order to check that the offset observed between the two datasets is not due to a flux calibration error in the GLEAM-X catalogue in the region around the QSO, we compared the GLEAM-X radio flux densities of the objects within 1$^\circ$ from the position of DES~J0209$-$56 to the ones reported in the Extragalactic GLEAM catalogue \citep{Hurley-Walker2017}. In particular, we only considered radio sources with a radio flux density $>40$~mJy in the GLEAM catalogue (since it is less sensitive compared to GLEAM-X) in the widest band available (170-231~MHz). Given a GLEAM RMS of $\sim6.5$~mJy~beam$^{-1}$ in this region, the flux density threshold corresponds to considering objects with a signal-to-noise $\gtrsim6$ in GLEAM. Specifically, we compared the integrated flux densities from the two catalogues, due to the different restoring beam size of the surveys. 
The differences between the GLEAM and GLEAM-X flux density measurements normalised by the GLEAM flux density are distributed as a Gaussian centred on 2\% and with a standard deviation corresponding to 8\%. From this analysis it is clear that the offset observed in the radio spectrum of DES~J0209$-$56 cannot be explained by the flux scale in the GLEAM-X survey, and therefore, it is likely associated with the observed radio emission of the object. 

A potential explanation for the observed difference is variability in the observed radio flux of DES~J0209$-$56. Indeed the GLEAM-X (observed between September 28 and October 25, 2020) and the ATCA (observed on June 24 and 25, 2023) measurements were taken about 2.7 years apart (or $\sim$5~months in the rest frame). However, we also note that the rest of the radio data points available around $\sim1$~GHz are consistent with the power law derived from the ATCA observations, despite being taken in different years: SUMSS between October 2000 and November 2001, RACS-low in May 2019, RACS-mid in January 2021, and EMU in December 2022. The consistency of all these data points with the extrapolation from the ATCA measurements suggests that variability, if present, only plays a marginal role in the observed spectral shape. At the same time, this could also mean that, for this particular source, variability is stronger at lower frequencies, differently from the general radio quasar population (e.g. \citealt{Sotnikova2024}). 

Another potential explanation for the offset observed in the radio spectrum of DES~J0209$-$56 is that different components of the jets are dominating the observed emission at low and high frequencies, similarly to the source PSO~J0202$-$17 described above. In this case both components have a flat emission ($\alpha_{\rm r}<0.5$), suggesting that they are produced in very compact, self-absorbed regions of the jet.
In any case, dedicated simultaneous observations over a wider range of frequencies (including at $\sim$300-700~MHz, in the observed frame) are needed in order to understand whether the difference between low- and high-frequency data is due to a complex shape of the radio spectrum or to variability.

From an X-ray point of view, the best-fit photon index derived from the \textit{Swift}-XRT observation is flat ($\Gamma_{\rm X}=1.38\pm0.46$; errors at 68\% confidence level). This value is consistent with the one derived by \cite{Wolf2024} using an independent dataset ($\Gamma_{\rm X}=1.46^{+0.44}_{-0.42}$; errors at 68\% confidence level), albeit both measurements have large uncertainties.
Using the spectral indices estimated from the best-fit radio, optical/NIR ($\alpha_{\rm o}^{\nu}=0.30\pm0.04$) and X-ray spectra together with the observed flux measurements closest to the rest-values that define each parameter, we derived $R=830\pm90$ and $\tilde{\alpha}_{\rm ox}=1.22\pm0.03$. In Fig. \ref{fig:SED}, it is clear that the X-ray emission observed in DES~J0209$-$56 is systematically above the one expected from the X-ray corona only, derived from \cite{Lusso2016}. Therefore, the strong X-ray and radio emission, compared to the optical one, of DES~J0209$-$56 suggest that this source is a blazar (consistent with the classification of \citealt{Wolf2024}), one of the most distant currently known in this class \citep[e.g.][]{Belladitta2020,Caccianiga2024,Banados2024}. At the same time, X-ray observations as well as a better sampling of the radio spectrum in the observed range $\sim200-800$~MHz are needed to better characterise the spectral shape of this source.

\subsection{PSO~J1011$-$01}

From a fit of the entire radio spectrum at $>800$~MHz in the observed frame, the spectral index obtained for PSO~J1011$-$01 is flat $\alpha_{\rm r} = 0.40\pm0.03$, which, again could indicate that the radio emission is dominated by relativistically boosted radiation. Even though this source is not reported in any radio catalogue at low frequency, there is a S/N$\sim$3.5 radio signal in the TGSS image at 150~MHz (S$_{\rm 150MHz}=7.4\pm2.1$~mJy). Interestingly, this data point is below the flux density expected from the extrapolation of the best-fit at higher frequencies, potentially implying a flattening of the radio spectrum at low frequencies, as also seen in other high-$z$ RL QSOs \citep[e.g.][]{Shao2022,Gloudemans2023}. However, as mentioned before for DES~J0209$-$56, this discrepancy could also be due to variability or the presence of another component in the radio spectrum.
A simultaneous sampling of the $\nu\sim100-600$~MHz range is needed to fully constrain the spectrum of this RL QSO. Using the best-fit radio and optical spectral indices ($\alpha_{\rm r} = 0.40\pm0.03$ and $\alpha_{\rm o}^{\nu}=0.27\pm0.03$) as well as the flux measurements closer to the desired rest-frame frequencies (RACS-low and W1 at 3.4~\textmu m), we obtained a radio-loudness parameter of ${\rm R} = 380\pm70$.

While the overall radio properties make PSO~J1011$-$01 a good blazar candidate, the non-detection in the X-rays suggests otherwise. 
Based on this \textit{Swift}-XRT non-detection, we can set a limit on the X-ray flux of PSO~J1011$-$01. Assuming an upper limit on the photon count in the 0.3--10~keV energy band of 3.2 photons in 10$''$ (90\% confidence interval for the measured background), the count rate corrected for the point-spread function of \textit{Swift}-XRT \citep[see][]{Moretti2005} is $\sim2.1\times10^{-4}$~counts~sec$^{-1}$. The corresponding intrinsic un-absorbed flux in the energy band 0.5--10~keV is $<8\times10^{-15}$~erg~cm$^{-2}$~sec$^{-1}$ and the associated luminosity is L$_{\rm2-10keV}<1.1\times10^{45}$ (assuming $\Gamma_{\rm X}=1.8$, as typical of radio-bright $z>4$ QSOs, e.g., \citealt{Zhu2019}). The $\tilde{\alpha}_{\rm ox}$ value is $>1.33$, which makes this source unlikely to be a blazar. However, based on the limits derived by \cite{Ighina2019}, the value measured does not completely rule out this possibility. Indeed, examples of blazars sources with a strong, compact and flat radio emission have been found to show a faint X-ray emission (see, e.g., \citealt{Belladitta2019,Gabany2023} and Fig. \ref{fig:R_aox}). Moreover, the non-detection could also be due to intrinsic variability of the source \citep[e.g.][]{Li2021}. 

\section{Comparison with high-$z$ RL QSO\lowercase{s}}
\begin{figure}
        \includegraphics[width=\hsize]{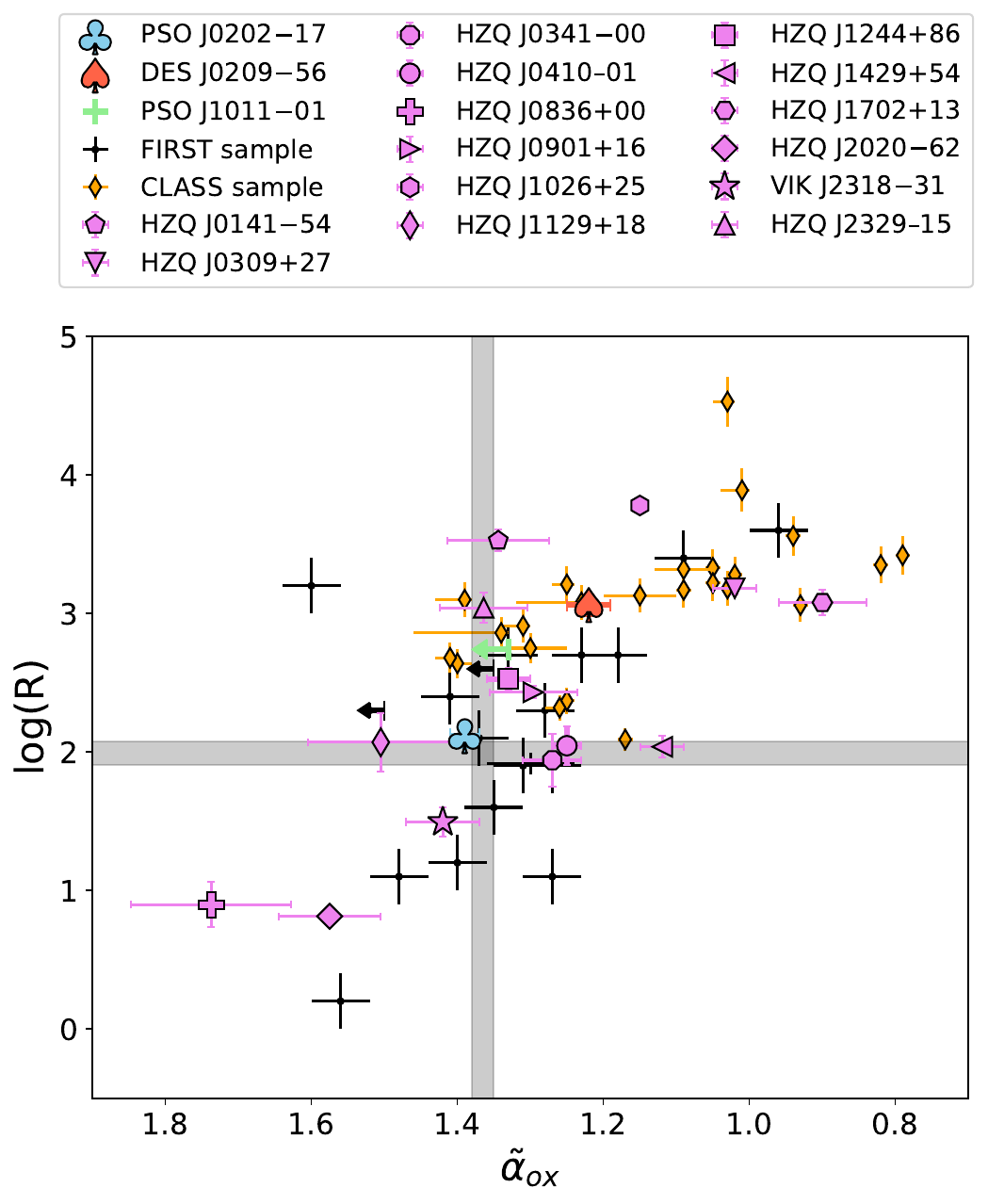}
    \caption{Radio loudness as a function of the $\tilde{\alpha}_{\rm ox}$ parameter for the objects presented in this work (blue circle, red square, and green upper limit) as well as for a complete radio-selected sample of $z>4$ blazars (\citealt{Caccianiga2019,Ighina2019}) and $z>4$ radio QSOs from the literature (\citealt{Caccianiga2024}, black points, and from single studies, pink symbols; see Table \ref{tab:aox_R_lit}). The shaded grey regions at $\tilde{\alpha}_{\rm ox}\sim1.35$ and log(R)~$\sim2$ divide the parameter space in two main regions: X-ray-, radio-bright objects (top-right corner, likely blazars) and X-ray-, radio-faint objects (bottom-left corner, likely hosting mis-aligned jets, i.e., classic RL QSOs).}
    \label{fig:R_aox}
\end{figure}

In this section we compare the multi-wavelength properties of the three $z\sim5.6$ RL QSOs described above to the other high-$z$ QSOs from the literature. In Fig. \ref{fig:R_aox}, we compare the radio loudness as a function of the $\tilde{\alpha}_{\rm ox}$ parameter for high-$z$ RL QSOs with information both in the radio and in the X-ray band. These two parameters quantify the intensity of the radio (R) and X-ray ($\tilde{\alpha}_{\rm ox}$) emission (both dominated by the radiation produced within the relativistic jets in blazars) with respect to the optical/UV one, dominated by the accretion disc in high-$z$ QSOs. 
As comparison, we considered two well-defined, radio-selected samples of high-$z$ QSOs: the first is composed by the $z>4$ QSOs detected in the FIRST radio survey with X-ray information from archives (described in \citealt{Caccianiga2024}, FIRST sample; black points); the second sample, CLASS sample, is a complete sample of S$_{\rm 5GHz}>30$~mJy sources with a flat radio spectral index ($\alpha_{\rm r}<0.5$; i.e. mostly blazars) at $z>4$ built by \cite{Caccianiga2019} and analysed in the X-rays by \cite{Ighina2019}. 
Finally, we also considered single $z>5$ radio QSOs discussed in dedicated works from the literature. The R and $\tilde{\alpha}_{\rm ox}$ values used in the plot for these $z>5$ radio QSOs (pink data points in Fig. \ref{fig:R_aox}) are reported in Table \ref{tab:aox_R_lit} together with the corresponding references.

From the plot, a clear trend is visible: sources bright in the X-rays (small $\tilde{\alpha}_{\rm ox}$ values) are also, on average, bright in the radio band (large R values). In particular, the parameter space can be divided in two main regions: sources with log(R)~$\lesssim2$ and $\tilde{\alpha}_{\rm ox}\gtrsim1.35$, where we can expect to find mis-aligned jets, and sources with log(R)~$\gtrsim2$ and $\tilde{\alpha}_{\rm ox}\lesssim1.35$, where we expect to find mostly blazars, for which both the radio and X-ray emission is relativistically boosted (see also discussion in \citealt{Caccianiga2024}). Indeed, most of the sources from the CLASS sample, selected to be radio powerful, are also X-ray powerful ($\tilde{\alpha}_{\rm ox}\lesssim1.35$), confirming their blazar nature. Same conclusions were also drawn by dedicated studies (including multi-frequency as well as VLBI data) on single objects in this region of the parameter space (e.g. HZQ~J0309+27 and HZQ~J1702+13; \citealt{Belladitta2020,Spingola2020,Khorunzhev2021,An2023,Liu2024}). At the same time, dedicated analysis on X-ray and radio-faint objects (log(R)~$\lesssim2$ and $\tilde{\alpha}_{\rm ox}\gtrsim1.35$) confirmed that the relativistic jets in these QSOs are mis-aligned with respect to our line of sight (e.g. HZQ~J0836+00 and VIK~J2318$-$31; \citealt{Frey2005,Wolf2021,Zhang2022,Ighina2024}).
However, when it comes to sources relatively radio- and/or X-ray-bright (i.e. log(R)~$\sim2$ and/or $\tilde{\alpha}_{\rm ox}\sim1.35$, shaded grey regions in Fig. \ref{fig:R_aox}), a blazar-non-blazar classification is not straightforward. Indeed, there are examples of high-$z$ objects that, despite having similar X-ray-to-optical luminosities and large radio luminosities, can either present blazar-like properties (e.g. HZQ~J0141$-$54; \citealt{Belladitta2019,Gabany2023}) or not (e.g. HZQ~J2329$-$15; \citealt{Banados2018b,Momjian2018,Connor2021}). 
Similarly, detailed analysis on X-ray powerful sources with R~$\sim100$ found that these sources can also be mis-aligned RL QSOs \citep[e.g. HZQ~J0341$-$00 and HZQ~J1429+54;][]{Frey2011,Medvedev2020,Medvedev2021,Zuo2024}. 

One of the difficulties in having a reliable classification of high-$z$ QSOs is also due to the presence of an evolution in their X-ray emission (e.g. \citealt{Zhu2019}), with high-$z$ RL QSOs being, on average, a factor of $\sim2$ brighter compared to similar objects at low redshift (e.g. \citealt{Ighina2019}). The origin of this enhancement is not clear yet, but one possibility is that it is related to the IC/CMB interaction, which, beside increasing the overall observed X-ray emission, is also expected to quench the radiation produced by the most extended regions of relativistic jets \citep[e.g.][]{Ghisellini2014b,Ghisellini2015,Afonso2015}.

Interestingly, in addition to an increase of the X-ray luminosity in high-$z$ RL QSOs, \cite{Zuo2024} also find hints of evolution of their X-ray spectral shape, with $z>4$ objects having, on average, steeper photon index values compared to their lower redshift counterparts and similar results have also been found when studying the most optically luminous radio-quiet QSOs at $z>6$ \citep[see][]{Zappacosta2023}. In this work we found two objects with a flat best-fit photon index ($\Gamma_{\rm X}\sim1.4$), albeit with quite large uncertainties, for an object that is likely a blazar (DES~J0209$-$56) and one which is potentially slightly mis-aligned (PSO~J0202$-$17). However, as mentioned also in \cite{Zuo2024}, the low photon index values might also be caused by a selection bias related to the radio-bright selection criteria. This would imply that the evolution of the photon index is due to different properties of the X-ray corona in high-$z$ QSOs, whose emission is only visible for radio-faint, mis-aligned RL QSOs. The X-ray analysis of well-defined samples (either radio- or optically selected) of RL QSOs at $z>5$ (i.e. where redshift-dependent evolutions are stronger; e.g. \citealt{Wu2013}) is needed in order to better constrain the observed evolution, both in terms of intensity as well as spectral shape, and understand its origin, whether it is related to the radiation produced within the jet \citep[e.g.][]{Ighina2021b} or the X-ray corona \citep[e.g.][]{Zhu2020}.

\section{Constraining the SMBH properties}
\label{sec:mass_BH}

In this section we use dedicated spectroscopic observations and public photometric optical/NIR observations of the sources discussed in this work in order to constrain the properties of the SMBH hosted at their centres. 
In particular, we use two independent approaches: analysis of the CIV broad-emission line (single epoch, SE, method) and the fit to the broad-band photometric measurements with an accretion disc (AD) model.

\subsection{Mass and accretion rate from the CIV emission line}
As described in Sect. \ref{sec:opt_data}, in the specific case of PSO~J1011$-$01 we obtained dedicated spectroscopic observations with LBT/LUCI aimed at covering the CIV broad emission line ($\sim$1~\textmu m in the observed frame for $z\sim5.5$). 

The use of this specific emission line is due to the fact that other broad-emission lines normally used in this type of analysis at high redshift are not accessible from ground-based telescope, or, if they are, they are heavily affected by telluric absorption (e.g. the MgII would be observed at $\sim$1.8~\textmu m at $z\sim5.5$). Following the analysis of the LUCI spectrum (Fig. \ref{fig:CIV_line}) described in Sect. \ref{sec:opt_data} and whose results are reported in Table \ref{tab:CIV_line}, we used the best-fit Gaussian function describing the line in order to estimate the mass of the central BH. While broad emission lines are normally described by more complex functions (e.g. \citealt{Richards2011,Coatman2017}), in this case we limited the analysis to a single Gaussian due to the low S/N of the spectrum and the lack of multiple components. 
In order to estimate the mass of the central SMBH in PSO~J1011$-$01, we used the following formula from \cite{Vestergaard2006}:

\begin{equation}
 \begin{aligned}
    {\rm M_{\rm BH}} = 10^{6.66} \times \left( \frac{\rm FWHM(CIV)}{\rm 10^3~km~sec^{-1}}\right)^2 \times \left(\frac{\lambda L_{1350\text{\normalfont\AA}}}{\rm 10^{44}~erg~sec^{-1}} \right)^{0.53} {\rm M_\odot},
\end{aligned}
    \label{eq:BHM_civ}
\end{equation}
where the $\lambda L_{1350\text{\normalfont\AA}}$ luminosity was computed from the photometric data points and spectral index. Using this relation, we find $\textrm{M}_{\rm BH}=2.2\pm0.5\times10^{9}~{\rm M_\odot}$, where the uncertainty only takes the statistical errors of the fit into account. Including the dispersion of the scaling relation reported in eq. \ref{eq:BHM_civ}, ${\sim} 0.36$~dex \citep{Vestergaard2006}, the corresponding best-fit value and uncertainty becomes: $\textrm{M}_{\rm BH}=2.2_{-1.2}^{+2.7}\times10^{9}~{\rm M_\odot}$.

As mentioned before, however, the CIV line can be affected by the presence of blue- or red-shifted components (e.g. \citealt{Sun2018}), associated with winds or other non-virialised regions (e.g. \citealt{Vietri2018}). Indeed, the potential shift between the CIV and the Ly$\alpha$ redshift estimates could indicate that also in the case of PSO~J1011$-$01 the CIV line is blue-shifted, with a velocity shift of $\sim1200$~km~sec$^{-1}$ in order to match the two redshift measurements. For this reason we also estimate the mass of the central SMBH in PSO~J1011$-$01 by considering the typical blue-shifts measured in $z\gtrsim6$ QSOs ($\Delta {\rm v}\sim2200$~km~sec$^{-1}$). In particular, by adopting the correction to the full width half maximum (FWHM) of the CIV line given by \cite{Coatman2017}, FWHM$_{\rm corr}$~=~FWHM~/~(0.41~$\times$~$\Delta$v~+~0.61), we found FWHM$_{\rm corr}=4150\pm390$. Using eq. \ref{eq:BHM_civ}, the corresponding mass is $\textrm{M}^{\rm corr}_{\rm BH}=1.0\pm0.3\times10^{9}~{\rm M_\odot}$, where the error is purely statistical. By including the scatter of the scaling relation, reduced to ${\sim} 0.20$~dex after the correction (see \citealt{Coatman2017}), the final estimate is $\textrm{M}^{\rm corr}_{\rm BH}=1.0_{-0.5}^{+0.6}\times10^{9}~{\rm M_\odot}$.


Starting from the $\lambda L_{1350\text{\normalfont\AA}}$ luminosity derived from the photometric data points of PSO~J1011$-$01 and assuming the bolometric correction derived by \cite{Shen2008}, $\rm K_{\rm bol}=3.81\pm1.26$, we derived the bolometric luminosity: $\rm L_{\rm bol}=4.7\pm1.7 \times 10^{46}$~erg~sec$^{-1}$. The corresponding Eddington ratio is $\rm \lambda_{\rm Edd}=L_{bol}/L_{Edd}=0.16\pm0.07$. By adding in quadrature the uncertainty on the scaling relation used to compute the mass of the central BH (eq. \ref{eq:BHM_civ}), the final uncertainty on the accretion rate is: $\rm \lambda_{\rm Edd}=0.16_{-0.11}^{+0.21}$. Similarly, by considering the value of the BH mass estimated by applying the correction derived from the typical blue-shift observed in high-$z$ QSOs, the final accretion rate is $\rm \lambda_{\rm Edd}=0.37_{-0.21}^{+0.27}$ (including the dispersion of the scaling relation).

\subsection{Mass and accretion rate from disc modelling}

In most high-$z$ QSOs the rest-frame optical-UV emission is dominated by the radiation produced from the accretion disc. Therefore, we can use the photometric data points in this band (observed optical/NIR) in order to constrain the properties of the accretion process and of the corresponding SMBH \citep[e.g.][]{Campitiello2018,Campitiello2020}. In particular, by adopting the \cite{Shakura1973} accretion disc model (SS73), we performed a fit to the available observations of all the three sources discussed in this work using the code reported in \cite{Rigamonti2024} (see also \citealt{Sbarrato2021,Belladitta2022} for similar applications).
The SS73 model assumes that the optical/UV continuum emission of the QSOs is produced by an optically thick, geometrically thin accretion disc composed by rings that emit as a black body with different temperatures, depending on the distance from the central SMBH (see, e.g., \citealt{Calderone2013} for a detailed description).

\begin{figure}
        \includegraphics[width=\hsize]{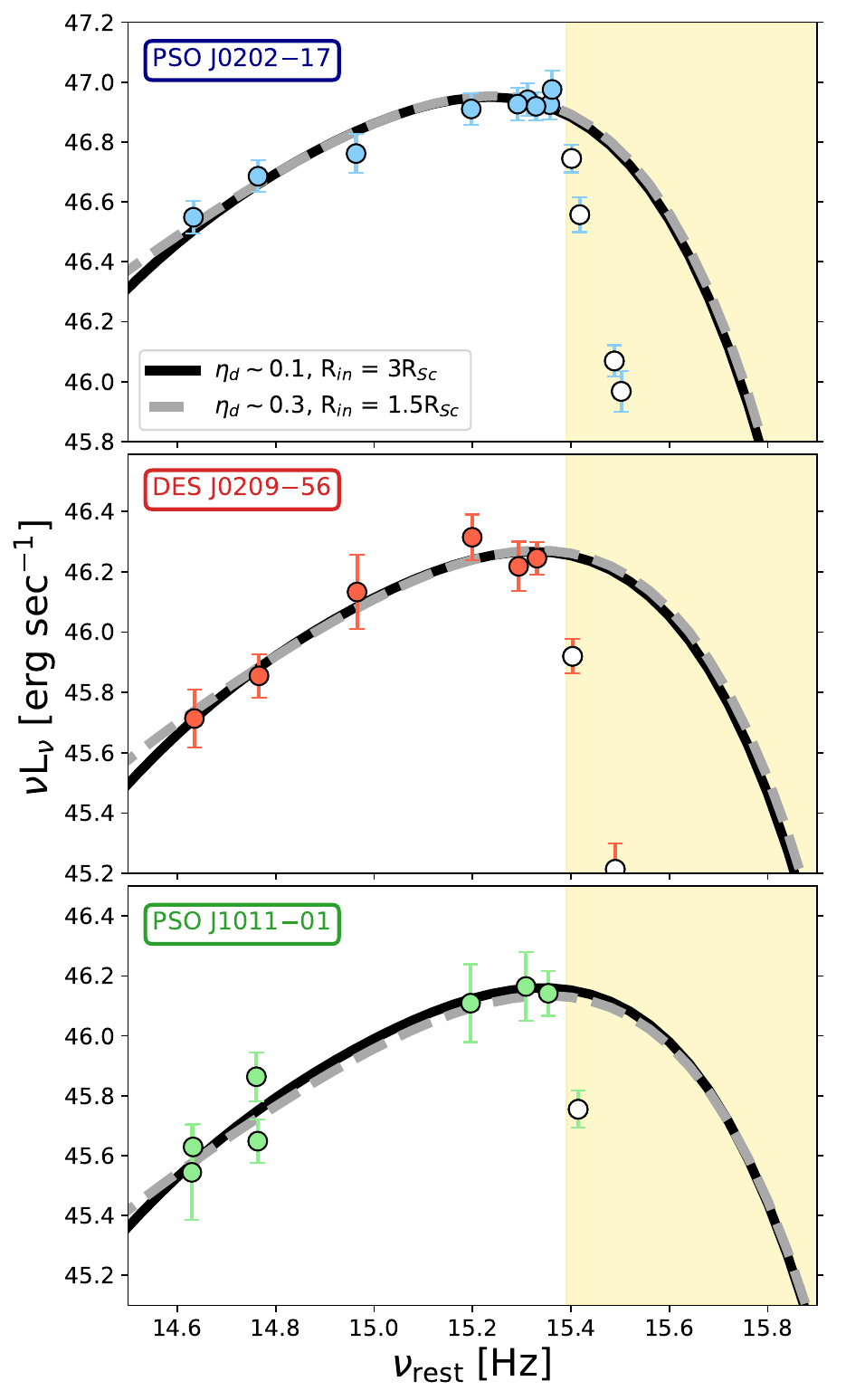}

    \caption{Rest-frame optical-UV SED of the three sources discussed in this work. The best-fit AD models derived for each target are reported as a black solid line ($\eta_{\rm d}\sim0.1$) and a grey dashed line ($\eta_{\rm d}\sim0.3$). During the fit, we only considered photometric data points (filled circles) not affected by IGM absorption (shaded area in yellow) and we did not use data affected, even marginally, by such absorption (white circles).}
    \label{fig:AD_fit}
\end{figure}

\begin{table*}
\caption{Best-fit values of the M$_{\rm BH}$ and $\lambda_{\rm Edd}$ (in the case $\eta_{\rm d}\sim0.1$ and $\eta_{\rm d}\sim0.3$) obtained from the AD modelling of the photometric measurements available for the $z>5.5$ RL QSOs discussed here.}
\centering
\begin{tabular}{lcccccccccccccc}
\\
\hline
\hline
Name & $z$ & M$_{\rm BH}$ & $\lambda_{\rm Edd}$ & M$_{\rm BH}$ & $\lambda_{\rm Edd}$ & M$_{\rm BH}$ & $\lambda_{\rm Edd}$ & M$_{\rm BH}$ & $\lambda_{\rm Edd}$ \\
     &     & (10$^9$~M$_\odot$) & & (10$^9$~M$_\odot$) & & (10$^9$~M$_\odot$ )& & (10$^9$~M$_\odot$) & &  \\

\hline
 && \multicolumn{2}{c}{$\eta_{\rm d}\sim0.1$} & \multicolumn{2}{c}{$\eta_{\rm d}\sim0.3$} & \multicolumn{2}{c}{SE estimate}  & \multicolumn{2}{c}{SE estimate corr.} \\
 \hline
PSO~J0202$-$17  & 5.57 & 4.9$^{+0.7}_{-0.7}$ & 0.19$^{+0.04}_{-0.04}$ & 9.6$^{+1.3}_{-1.3}$ & 0.15$^{+0.03}_{-0.03}$ & -- & -- & -- & -- \\
DES~J0209$-$56 & 5.61 & 1.5$^{+0.4}_{-0.4}$ & 0.13$^{+0.05}_{-0.04}$ & 2.8$^{+0.6}_{-0.9}$ & 0.10$^{+0.04}_{-0.03}$ & -- & -- & -- & -- \\
PSO~J1011$-$01  & 5.56 & 1.2$^{+0.3}_{-0.4}$ & 0.13$^{+0.04}_{-0.04}$ & 2.2$^{+0.6}_{-0.7}$ & 0.10$^{+0.04}_{-0.03}$ & $2.2_{-1.2}^{+2.7}$ & $0.16_{-0.11}^{+0.21}$ & $1.0_{-0.5}^{+0.6}$ & $0.37_{-0.21}^{+0.27}$\\
\hline
\hline

\end{tabular}

\tablefoot{The errors correspond to the 16$^{\rm th}$ and 84$^{\rm th}$ percentile values obtained using a MCMC algorithm. For PSO~J1011$-$01 only, we also include the estimates based on the CIV line, both the un-corrected and the corrected (for the typical blue-shift) values. In both cases the uncertainties take the scatter of the scaling relation adopted into account.}
\label{tab:AD_masses}
\end{table*}

In this specific case we set the outer radius of the disc to the self-gravitation radius of the disc (see, eq. 2 in \citealt{Netzer2015}), which, for a BH with ${\rm M}_{\rm BH}\sim10^9_{\odot}$ and a radiative efficiency $\eta_{\rm d}\sim0.1$, is R$_{\rm out} \sim 4\times10^2$~R$_{\rm S}$, where R$_{\rm S}$ is the Schwarzschild radius.\footnote{R$_{\rm S}=2G{\rm M_{\rm BH}}/c^2$, with $G$ the gravitation constant and $c$ the speed of light.} We note that the actual value of this parameter does not significantly affect the shape of the SED in the wavelength range considered here. The choice of the inner radius, instead, is more important, as it is closely related to the efficiency of the accretion process: for  a BH rotating close to its maximum value we have $\eta\sim0.3$ and $\rm R_{\rm in}\sim1.5R_{\rm S}$, while for a non-rotating BH $\eta\sim0.1$ and $\rm R_{\rm in}\sim3R_{\rm S}$ (e.g. \citealt{Calderone2013}), where $\eta$ is the fraction of the overall accreting material that is converted into energy. 
We note that, for simplicity, in this section we assumed that all the energy available from the accretion process goes into radiation (i.e. $\eta=\eta_{\rm d}$). However, as mentioned before, it is also possible that a fraction of this energy is also responsible for the magnetic field enhancement and, as a consequence, the launch of relativistic jets (e.g. \citealt{Jolley2008}). Even though a value of $\eta_{\rm d}\sim0.1$ is normally assumed in this type of modelling (\citealt{Sbarrato2015,Belladitta2022}), here we considered both the $\eta_{\rm d}\sim0.1$ and the $\eta_{\rm d}\sim0.3$ cases, since QSOs hosting powerful relativistic jets are likely associated with fast spinning BHs \citep[e.g.][]{Chen2021}, in order to show how much this assumption affects the results. 
Our simplified model does not take relativistic corrections (e.g. gravitational redshift, light bending and self-irradiation; \citealt{Li2005}) into account, which can significantly affect the spectral shape and intensity observed. As a reference, by including general relativity corrections to the black-body spectrum of the innermost regions of the disc, our simplified model assuming $\eta_{\rm d}\sim0.1$ is a good approximation of a rotating BH with spin $a=0.71$ (see, e.g., fig. A2 in \citealt{Calderone2013}).
Finally, for the inclination of the disc with respect to our line of sight, we considered $\theta_{\rm v}$ to be $\sim0^\circ$ in the case of DES~J0209$-$56 (likely a blazar) and $\sim20^\circ$ in the case of PSO~J0202$-$17 and PSO~J1011$-$01 ( type 1 QSOs with jets likely mis-aligned with respect to our line of sight), based on the discussion of Sect. \ref{sec:ind_obj}. We note that the corresponding normalisation factor of the optical SED only varies from 2 (for $\theta_{\rm v}\sim0^\circ$) to 1.9 (for $\theta_{\rm v}\sim20^\circ$; see Sect. A3 in \citealt{Calderone2013}) and the resulting best-fit $M_{\rm BH}$ and $\lambda_{\rm Edd}$ vary only by a few percent.

To perform the fit with the model described above, we considered the photometric measurements available from public surveys, all corrected for galactic extinction (see Table \ref{tab:photom_data} and coloured data points in Fig. \ref{fig:AD_fit}). During the fit, we excluded all the filters that are affected, even marginally, by the IGM absorption (i.e. shortwards of the Ly$\alpha$ line; white data points in Fig. \ref{fig:AD_fit}). Moreover, in the case of PSO~J0202$-$17, we also excluded the $W3$ filter, since, as shown in Fig. \ref{fig:SED}, it is likely tracing the emission produced by another components (likely the dusty torus). Since the optical/NIR measurements come from several observations performed at different epochs, we also included an additional 10\% in their error, added in quadrature, in order to account for potential variability \citep[e.g. ][]{MacLeod2010,Dexter2011}.


Figure \ref{fig:AD_fit} shows the best-fit model obtained for the three $z>5.5$ RL QSOs, assuming $\eta_{\rm d}\sim0.1$ (solid black line) and $\eta_{\rm d}\sim0.3$ (dashed grey line). As clear from the plots, both models can well reproduce the observed emission and it is not possible to distinguish between the two based on the goodness of the fit only. We report in Table \ref{tab:AD_masses} the best-fit values obtained for M$_{\rm BH}$ and $\lambda_{\rm Edd}$ together with the corresponding 16$^{\rm th}$ and 84$^{\rm th}$ percentile values obtained using the \texttt{emcee}, \citealt{Foreman2013}, Markov chain Monte Carlo (MCMC) algorithm with a uniform prior distribution. 
In general, the best-fit M$_{\rm BH}$ change by a factor $\sim2$ based on the choice $\eta_{\rm d}$, while the $\lambda_{\rm Edd}$ values are more stable. This is because increasing $\eta_{\rm d}$, which within our model assumptions implies a decrease in R$_{\rm in}$, has a similar effect of decreasing the BH mass. Indeed, in both cases, having more radiation from the innermost hot part of the disc included in the overall emission causes a shift towards bluer frequencies (e.g. \citealt{Calderone2013}). 
Therefore, a factor of around two is a more realistic estimate for the uncertainty associated with the M$_{\rm BH}$ derived from our AD model.
In the specific case of PSO~J1011$-$01, the values obtained with the AD modelling are in good agreement with the independent measurement based on the CIV broad-emission line (including the blue-shift corrected estimate), indicating these three $z>5.5$ RL QSOs are hosting SMBH with a mass M$_{\rm BH}\sim1-10\times10^9$~M$_\odot$ accreting at a rate $\lambda_{\rm Edd}\sim0.1-0.4$. These values are consistent with the ones derived for other high-$z$ QSOs, both radio-quiet (e.g. \citealt{Mazzucchelli2023,Lai2024}) and RL (e.g. \citealt{Diana2022,Belladitta2022}), and are likely a consequence of the optical selection.

\section{Summary and conclusions}

In this work we have presented three new $z\sim5.6$ radio-powerful QSOs discovered from the combination of the RACS-low survey \citep{McConnell2020} with the Pan-STARRS \citep{Chambers2016} and DES \citep{Abbott2021} surveys. 
Their high-$z$ nature was confirmed with spectroscopic observations from the AAT/KOALA, Gemini-S/GMOS and VLT/FORS2 telescopes, making them some of the most radio-bright ($\rm R>100$) RL QSOs currently known at $z>5.5$ \citep[e.g.][]{Banados2021,Ighina2021a,Gloudemans2021}. For all three sources, we also obtained dedicated observations in the radio (ATCA) and in the X-ray bands, providing complementary information with respect to data available from radio surveys. Based on the analysis of their multi-wavelength SEDs, the three QSOs present a wide range of properties in the different electromagnetic bands in terms of intensity and spectral shape. 

PSO~J0202$-$17 presents a strong X-ray luminosity with a photon index $\Gamma_{\rm X}<1.9$. However, when compared to the optical luminosity, which is also strong, its $\tilde{\alpha}_{\rm ox}=1.39\pm0.02$ parameter is consistent with the one expected from the X-ray corona emission \citep[e.g.][]{Vignali2005}. Therefore, in this specific case, the emission produced by the relativistic jets and the X-ray corona are comparable. In the radio band, the spectrum of PSO~J0202$-$17, which consists of two components, is more complex than for other $z>5$ radio QSOs \citep[e.g.][]{Shao2022,Gloudemans2022}. Indeed, at low frequencies ($<300$~MHz), the simultaneous data points observed in GLEAM-X can be well described by a simple power law with spectral index $\alpha_{\rm low}\sim0.7$, while higher-frequency ($>2$~GHz) simultaneous ATCA observations indicate the presence of a peaked component in the radio emission at $\sim6$~GHz in the observed frame ($\sim40$~GHz in the rest frame). Further radio observations at different angular scales are needed to better understand the origin and the properties of the radio jets in PSO~J0202$-$17.

DES~J0209$-$56 shows strong radio and X-ray emission, indicating that the observed radiation in these bands is  dominated by the Doppler-boosted radiation produced within the jets, that is, this source is a blazar (e.g. \citealt{Ghisellini2015c}). This discovery corresponds to one of the only four blazars currently known at $z>5.5$ (\citealt{Belladitta2020,Caccianiga2024,Banados2024}). Deeper X-ray observations as well as simultaneous radio observations over a wide range of frequencies (0.1-10~GHz) are needed in order to firmly constrain the spectral shape of DES~J0209$-$56 in these bands.

Finally, the source PSO~J1011$-$01 presents a flat radio emission ($\alpha_{\rm r}\sim0.4$) with a possible flattening at low frequencies, as suggested by the 150-MHz measurement from TGSS. Simultaneous observations at low frequencies ($<600$~MHz; e.g. with the next data release of the MWA-GLEAM-X survey) are needed to constrain the radio spectral shape of this source. At the same time, the target is not detected in 30~ksec of \textit{Swift}-XRT observations, implying an upper limit on the intrinsic X-ray flux of $\lesssim0.8\times10^{-14}$~erg~sec$^{-1}$~cm$^{-2}$ (assuming a power-law emission with $\Gamma_{\rm X}=1.8$). While the non-detection in the X-rays suggests that the radiation produced by the relativistic jets is not strongly boosted, the upper-limit derived for PSO~J1011$-$01 could still be consistent with a blazar classification (e.g. \citealt{Ighina2019}). Moreover, intrinsic variability effects might also impact the observed flux. 

From the analysis of the UV/optical rest-frame emission of the three $z\sim5.6$ RL QSOs described in this work, we were able to constrain the properties of their SMBHs. In particular, based on the modelling of their photometric data points with a SS73 AD model as well as the analysis of the CIV line in the case of PSO~J1011$-$01, these sources host SMBHs with a mass of $\rm M_{\rm BH}\sim1-10 \times 10^9$~M$_\odot$ accreting at a fraction of their Eddington limit ($\lambda_{\rm Edd}\sim0.1-0.4$). These values are consistent with the ones derived for other high-$z$ QSOs selected from the same optical surveys using a similar method \citep[e.g.][]{Shen2019,Mazzucchelli2023}, and they challenge our current models for seed BH formation and evolution \citep[e.g.][]{Volonteri2021}. However, in contrast to the majority of $z>5.5$ QSOs currently known, these three objects host radio-powerful relativistic jets whose expected theoretical accretion efficiency is $\eta\sim0.3$ \citep[as expected for fast rotating BHs, e.g.,][]{Thorne1974}. If all of the energy produced during the accretion process went into producing radiation (i.e. $\eta=\eta_{\rm d}$), it would make the presence of $\rm M_{\rm BH}\sim1-10 \times 10^9$~M$_\odot$ SMBHs hosted in these RL QSOs at $z\sim5.6$ even harder to reproduce with our current theoretical models. If the majority of the energy released during the accretion process is used to help the jet-launching mechanism by increasing the AD magnetic field ($\eta=\eta_{\rm d} + \eta_{\rm j}$, with $\eta_{\rm j}>>\eta_{\rm d}$, e.g. \citealt{Jolley2008}), the central BH hosted in an RL QSO could actually grow faster compared to one hosted in a similar radio-quiet source since it would need to accrete more matter in order to produce the same amount of optical luminosity (see, e.g., \citealt{Connor2024}).

The discovery and the multi-wavelength characterisation of more $z>5.5$ jetted systems is crucial to better understanding the evolution of relativistic jets in the early universe (e.g.; \citealt{Ighina2021b,Zuo2024}) as well as their impact on the growth of the first seed BHs \citep[e.g.][]{Jolley2008,Ghisellini2013}. Up-coming radio and optical/NIR wide-area surveys surveys such as EMU \citep{Norris2011,Norris2021}, the Vera C. Ruby Observatory \citep{Ivezic2019} and the EUCLID-wide survey \citep{Euclid2022} will allow us to significantly increase the number of high-$z$ jetted quasars, even well within the epoch of re-ionisation ($z>7$, see, e.g., \citealt{Ighina2023}).

\begin{acknowledgements}
We want to thank J. Afonso and C. Vignali for their useful comments and K. Ross for helping with the GLEAM-X data.
L.I. would like to thank all the staff working at the ATCA/Paul Wild Observatory and AAT/Siding Spring observatories for the great support and the amazing experience provided during the observations.
We thank the referee for their comments on this manuscript.\\ 

We acknowledge financial support from INAF under the project ``QSO jets in the early Universe'', Ricerca Fondamentale 2022 and under the project ``Testing the obscuration in the early Universe'', Ricerca Fondamentale 2023.\\
F.R. acknowledges the support from the Next Generation EU funds within the National Recovery and Resilience Plan (PNRR), Mission 4 - Education and Research, Component 2 - From Research to Business (M4C2), Investment Line 3.1 - Strengthening and creation of Research Infrastructures, Project IR0000012 – ``CTA+ - Cherenkov Telescope Array Plus''.\\
This scientific work uses data obtained from Inyarrimanha Ilgari Bundara / the Murchison Radio-astronomy Observatory. We acknowledge the Wajarri Yamaji People as the Traditional Owners and native title holders of the Observatory site. CSIRO’s ASKAP radio telescope is part of the Australia Telescope National Facility (\url{https://ror.org/05qajvd42}). Operation of ASKAP is funded by the Australian Government with support from the National Collaborative Research Infrastructure Strategy.\\
The Australia Telescope Compact Array is part of the Australia Telescope National Facility (\url{https://ror.org/05qajvd42}) which is funded by the Australian Government for operation as a National Facility managed by CSIRO. We acknowledge the Gomeroi people as the Traditional Owners of the Observatory site.\\
This project has received funding from the European Union's Horizon 2020 research and innovation programme under grant agreement No 101004719. This material reflects only the authors views and the Commission is not liable for any use that may be made of the information contained therein.\\
This work was supported by resources provided by the Pawsey Supercomputing Research Centre with funding from the Australian Government and the Government of Western Australia.\\
We acknowledge the support from the LBT-Italian Coordination Facility for the execution of the observations. This research used the facilities of the Italian Center for Astronomical Archive (IA2) operated by INAF at the Astronomical Observatory of Trieste.\\
We acknowledge the use of public data from the \textit{Swift} data archive.\\
This publication makes use of data products from the Wide-field Infrared Survey Explorer, which is a joint project of the University of California, Los Angeles, and the Jet Propulsion Laboratory/California Institute of Technology, funded by the National Aeronautics and Space Administration.
\end{acknowledgements}

%
\bibliographystyle{aa} 
\bibliography{referenze} 
%

\begin{appendix}
    \onecolumn
\section{Radio and X-ray properties of $z>5$ QSOs from the literature}
\label{app:lit_sample}

In this section we report the values of the $\tilde{\alpha}_{\rm ox}$ parameter and of the radio loudness for several $z>5$ radio QSOs analysed in the literature that are shown in Fig. \ref{fig:R_aox}.

\begin{table*}[h!]
\caption{Values of the $\tilde{\alpha}_{\rm ox}$ parameter and radio loudness of $z>5$ QSOs from the literature shown in Fig. \ref{fig:R_aox}. The references reported in this table are for the discovery, the $\tilde{\alpha}_{\rm ox}$ parameter and the radio loudness.} 
\centering
\begin{tabular}{lcccl}
\\
\hline
\hline
Name & $z$ & $\tilde{\alpha}_{\rm ox}$ & log(R) & References \\
\hline
\\
HZQ~J0141$-$54 & 5.000 & 1.34 $\pm$ 0.07 & 3.53 $\pm$ 0.08 & \cite{Belladitta2019}\\
HZQ~J0341$-$00 & 5.68 & 1.27 $\pm$ 0.04 & 1.94 $\pm$ 0.19 & \cite{Banados2015,Zuo2024}\\
HZQ~J0309+27 & 6.100 & 1.02 $\pm$ 0.03 & 3.18 $\pm$ 0.04 & \cite{Belladitta2020,Spingola2020,Moretti2021}\\
HZQ~J0410$–$01 & 6.996 & 1.25 $\pm$ 0.02 & 2.05 $\pm$ 0.14 & \cite{Banados2024} \\ 
HZQ~J0836+00 & 5.81 & 1.74 $\pm$ 0.11 & 0.90 $\pm$ 0.16 & \cite{Fan2001,Wolf2021}\\
HZQ~J0901$+$16 & 5.63 & 1.30 $\pm$ 0.06 & 2.43 $\pm$ 0.04 & \cite{Banados2015,Caccianiga2024}\\
HZQ~J1026$+$25 & 5.25 & 1.15 $\pm$ 0.01 & 3.78 $\pm$ 0.03 & \cite{Sbarrato2012,Ighina2019}\\
HZQ~J1129+18 & 6.823 & 1.50 $\pm$ 0.10 & 2.07 $\pm$ 0.21 & \cite{Banados2021,Zuo2024}\\
HZQ~J1244+86 & 5.320 & 1.33 $\pm$ 0.03 & 2.53 $\pm$ 0.01 & \cite{Belladitta2023} \\
HZQ~J1429+54 & 6.184 & 1.12 $\pm$ 0.03 & 2.04 $\pm$ 0.08 & \cite{willott2005,Frey2011,Migliori2023}\\
HZQ~J1702+13 & 5.466 & 0.90 $\pm$ 0.06 & 3.08 $\pm$ 0.09 & \cite{Khorunzhev2021,An2023}\\
HZQ~J2020$-$62 & 5.718 & 1.57 $\pm$ 0.07 & 0.81 $\pm$ 0.06 & \cite{Wolf2024}\\
HZQ~J2318$-$31 & 6.443 & 1.42 $\pm$ 0.05 & 1.49 $\pm$ 0.10 & \cite{Decarli2018,Ighina2024}\\
HZQ~J2329$-$15 & 5.832 & 1.36 $\pm$ 0.06 & 3.04 $\pm$ 0.11 & \cite{Banados2018b,Rojas-Ruiz2021,Connor2021}\\

\hline
\hline

\end{tabular}

\label{tab:aox_R_lit}
\end{table*}
\end{appendix}



\end{document}